\documentclass[conference, onecolumn]{IEEEtran}
\usepackage{url}

\newcommand{\name}{\textsc{Veritas}\;}
\newcommand{\namecomma}{\textsc{Veritas}}

\usepackage{hyperref}

\usepackage{tikz}
\usepackage{amsmath}
\usepackage{amssymb}
\usepackage{mathtools} 
\usepackage{amsthm}
\usepackage{esvect} 
\usepackage{mathtools} 
\usepackage{bbm}
\usepackage{bbold}
\let\labelindent\relax
\usepackage{enumitem}

\usepackage{caption}
\usepackage{subcaption}

\newtheorem{theorem}{Theorem}

\usepackage{array}
\newcolumntype{L}[1]{>{\raggedright\let\newline\\\arraybackslash\hspace{0pt}}m{#1}}
\newcolumntype{C}[1]{>{\centering\let\newline\\\arraybackslash\hspace{0pt}}m{#1}}
\usepackage{multirow}
\usepackage{tabularx}
\usepackage{rotating}
\usepackage{adjustbox}
\newcolumntype{R}[2]{%
    >{\adjustbox{angle=#1,lap=\width-(#2)}\bgroup}%
    l%
    <{\egroup}%
}
\newcommand*\rot{\multicolumn{1}{R{25}{1em}}}%
\usepackage{xcolor,colortbl}

\usepackage{fancybox}

\usepackage{pifont}
\newcommand{\cmark}{\ding{51}}%
\newcommand{\xmark}{\ding{55}}%

\usepackage{nomencl} 
\makenomenclature
\usepackage{etoolbox}
\renewcommand\nomgroup[1]{%
  \item[\bfseries
  \ifstrequal{#1}{H}{Homomorphic Encryption}{%
  \ifstrequal{#1}{E}{Encodings}{%
  \ifstrequal{#1}{O}{Other}{}}}%
]}

\usepackage[n, advantage, operators, sets, adversary, landau, probability, notions, logic, ff, mm, primitives, events, complexity, oracles, asymptotics, keys]{cryptocode}
\usepackage{dashbox}

\createprocedureblock{procbox}{center,boxed}{}{}{}

\newcounter{scheme}
\createprocedureblock{scheme}{center,boxed}{}{$\textbf{Scheme~\thescheme}$:~}{mode=text,jot=3mm, bodylinesep=1mm, width=0.6\textwidth, colspace=-1cm}

\newcounter{expe}
\createprocedureblock{experiment}{center,boxed}{}{$\textbf{Experiment~\theexpe}\!:$~}{mode=text,jot=3mm, width=0.6\textwidth, colspace=-1cm}

\newcounter{oracle}
\createprocedureblock{oracle}{center,boxed}{}{$\textbf{Oracle~\theoracle}\!:$~}{mode=text,jot=3mm, width=0.6\textwidth, colspace=-1cm}

\newcommand{\descr}[1]{\smallskip \noindent \textbf{#1}}

\newcommand{\descrit}[1]{\smallskip \noindent \textit{#1}}
\newcommand{\etal}{\textit{et al.\,}}
\newcommand{\ie}{\textit{i.e.,\,}}
\newcommand{\eg}{\textit{e.g.,\,}}

\begin{document}

\title{Verifiable Encodings for Secure Homomorphic Analytics}

\author{
    \IEEEauthorblockN{Sylvain Chatel\IEEEauthorrefmark{1}\IEEEauthorrefmark{2}, Christian Knabenhans, Apostolos Pyrgelis\IEEEauthorrefmark{2}, Carmela Troncoso, and Jean-Pierre Hubaux\\ }
    \IEEEauthorblockA{Swiss Federal Institute of Technology in Lausanne (EPFL)\\
    }
    \IEEEauthorblockA{\IEEEauthorrefmark{1}Email: first.last@epfl.ch\\
    }
}

\IEEEoverridecommandlockouts
\makeatletter\def\@IEEEpubidpullup{3.5\baselineskip}\makeatother
\IEEEpubid{\parbox{\columnwidth}{
    Contact email: sylvain.chatel@epfl.ch\\
}
\hspace{\columnsep}\makebox[\columnwidth]{}}

\maketitle
\thispagestyle{plain}
\pagestyle{plain}

\begin{abstract}
Homomorphic encryption has become a practical solution for protecting the privacy of computations on sensitive data. However, existing homomorphic encryption pipelines do not guarantee the \textit{correctness} of the computation result in the presence of a malicious adversary. We propose two plaintext encodings compatible with state-of-the-art fully homomorphic encryption schemes that enable practical client-verification of homomorphic computations while supporting all the operations required for modern privacy-preserving analytics. Based on these encodings, we introduce \namecomma, a ready-to-use library for the verification of computations executed over encrypted data. \name is the first library that supports the verification of \textit{any} homomorphic operation. We demonstrate its practicality for various applications and, in particular, we show that it enables verifiability of homomorphic analytics with less than $3\times$ computation overhead compared to the homomorphic encryption baseline.
\end{abstract}

\renewcommand{\thefootnote}{\fnsymbol{footnote}}
\footnotetext[2]{Authors were affiliated with EPFL at the time of this work.}
\renewcommand{\thefootnote}{\arabic{footnote}}

\section{Introduction}\label{sec:intro}
Homomorphic encryption (HE) enables a computing server to perform operations on encrypted data without decrypting it. It has become an auspicious technique to support practical privacy-preserving computations in various scenarios, ranging from pure computation outsourcing~\cite{chen2018logistic,cheon2017privacy,kim2015private}, to distributed computation among multiple clients~\cite{chen2017batched,chen2019efficient,mouchet2021multiparty,sav2021poseidon,froelicher2023scalable}. Due to the wide range of operations they support and their efficiency, lattice-based HE schemes, \eg BFV~\cite{fan2012somewhat,halevi2019improved} and BGV~\cite{brakerski2014leveled,kim2021revisiting}, are becoming the basis for many privacy-preserving applications such as medical research analytics~\cite{cheon2017privacy,kim2015private,lu2015privacy,sav2022privacy}, private information retrieval~\cite{boneh2013private}, private set intersection~\cite{chen2017fast}, and machine learning (ML)~\cite{brutzkus2019low,chen2018logistic,graepel2012ml,lu2016using,sav2021poseidon}.

However, existing HE schemes~\cite{fan2012somewhat,halevi2019improved,brakerski2014leveled,kim2021revisiting} do not provide clients with any guarantees about the \emph{correctness} of the computations performed by the computing server. Thus, a malicious adversary can break the security and privacy of the HE computation without being detected~\cite{chatel2023pelta,viand2023verifiable}. For instance, clients can be fooled into accepting a wrongful computation result and potentially leak unintended information from the decryption~\cite{chenal2014key}. In current HE applications, the lack of computational integrity can lead to catastrophic consequences. For example, in medical applications, an adversary inducing a wrongful prediction might cause a misdiagnosis, and in machine learning an adversary able to inject backdoors during training can create vulnerabilities when the model is deployed~\cite{bagdasaryan2021blind,biggio2018wild,biggio2012poisoning}.

A trivial way for clients to check the correctness of the computation would be to recompute the result in plaintext. However, this is not always feasible, \eg in multi-client scenarios where not all the input data is available to the clients. To address this problem, researchers have proposed generic verifiable computation techniques, \eg\cite{bhadauria2020ligero++,bunz2018bulletproofs,goldwasser2015delegating,gennaro2010non,parno2013pinocchio,weng2021wolverine}, to check the integrity of server computations. These techniques, however, are hard to integrate with lattice-based HE. This is because the efficiency gains that make modern HE practical (\eg for medical research~\cite{cheon2017privacy,kim2015private,lu2015privacy} and ML~\cite{brutzkus2019low,chen2018logistic,graepel2012ml,lu2016using,sav2021poseidon,sav2022privacy}) stem from the use of (i) specific polynomial constructions and algebraic parameters, and (ii) non-linear operations (\eg relinearization and rotation) and parallel processing over encrypted data (Single Input, Multiple Data; SIMD). Generic verification techniques cannot cope with all these constraints efficiently.

To date, only a handful of verification works are tailored to homomorphically encrypted data. Fiore \etal~\cite{fiore2014efficiently} pioneered a solution based on homomorphic Message Authentication Codes (MACs) to verify encrypted computations. Their work focuses only on verifying quadratic functions over a constrained version of the BV scheme~\cite{brakerski2011fully}. Thus, it cannot support flexible parameterization (\eg large-degree polynomials) and complex operations (\eg rotations) required to implement modern HE-based applications. Subsequent works rely on generating a proof-of-correct computation using succinct non-interactive arguments of knowledge (SNARKs)~\cite{fiore2020boosting,bois2021flexible,ganesh2021rinocchio}. But, due to incompatibilities of SNARKs with the algebraic structure of recent HE schemes and their non-algebraic operations (\eg rounding) these works also do not support variable HE parameters~\cite{fiore2020boosting} and core HE operations~\cite{bois2021flexible,ganesh2021rinocchio} such as relinearization, rotation, and bootstrapping. Without support for flexible HE parameters and critical HE operations, these works have very limited application in practice.

\begin{table}[!t]
\caption{Comparison between \name and prior work w.r.t. the supported HE operations (linear, multiplicative, rotation, relinearization, bootstrapping) and parameterization.}
\label{tab:rw}
\centering
\fontsize{8}{7.2}\selectfont
\setlength{\tabcolsep}{4pt}
\begin{tabular}{C{1.1cm}|>{\centering\arraybackslash\columncolor[gray]{0.9}}C{.8cm}|C{.9cm}|>{\centering\arraybackslash\columncolor[gray]{0.9}}C{.8cm}|C{.8cm}|>{\centering\arraybackslash\columncolor[gray]{0.9}}C{1.2cm}|C{.7cm}}
 Scheme & Linear & Mult. depth & Relin. & Rot. & Bootstrap. & Flex. params. \\
\hline
\cite{fiore2014efficiently} & \cmark & 1 & \xmark & \xmark & \xmark & \xmark\\
\cite{fiore2020boosting} & \cmark & any & \xmark & \xmark & \xmark & \xmark\\
\cite{bois2021flexible,ganesh2021rinocchio} & \cmark & any & \xmark & \xmark & \xmark & \cmark\\
\textbf{Ours}& \cmark & any & \cmark & \cmark & \cmark & \cmark\\
\end{tabular} 
\end{table}

\descr{Our Contribution.} We provide the first practical solution that enables the verifiability of \textbf{all} the operations supported by state-of-the-art lattice-based HE schemes over the integers (see Table~\ref{tab:rw}). Therefore, our solution supports any existing privacy-preserving application that employs such HE schemes and it can protect its computation integrity against a malicious computing server oblivious of the decryption result and the verification outcome. Additionally, our solution does not require the input data to verify the computation which makes it applicable to settings with multiple trusted clients and a malicious server.

The key idea behind our solution is to shift the verification from the ciphertext domain, where the constraints imposed by the algebraic structure of HE ciphertexts limit the practicality of previous solutions, to the plaintext space. To this end, we design two new plaintext HE-encoders that, combined with HE schemes, instantiate homomorphic authenticators~\cite{catalano2013practical,chung2010improved,gennaro2013fully}, \ie efficient constructions to verify homomorphically-executed computations. These authenticators permit the verification of the computations by checking a set of challenge values that can be pre-computed without access to the original input data (\ie this pre-computation is an optimization and can be assigned to any entity that does not collude with the computing server). This makes our solution suitable for multi-party scenarios~\cite{chen2017batched,chen2019efficient,mouchet2021multiparty} and computationally constrained clients. Our first encoding is based on \textit{replication} and takes advantage of the batching encoder supported by modern HE to introduce error-detection and redundancy in the data. The second one achieves more compact authentications and requires less challenge values to be verified by encoding the data using a polynomial information-theoretic MAC. The two encoders achieve different tradeoffs depending on the volume of input data and the depth of the computation, hence, they can support various application settings.

Using our novel encodings, we design \namecomma, an open-source library that facilitates the conversion of existing HE computing pipelines, that assume an honest-but-curious computing server, into pipelines that can detect a malicious but rational server~\cite{aumann2007security}. We benchmark \name on native HE operations and evaluate its performance on five use cases: ride-hailing, genomic-data analysis, encrypted search, and machine-learning training and inference, where computation integrity is crucial. Our results demonstrate that \name provides verification capabilities that were out of reach using prior work, at minimal costs for the client and the server. For instance, our solution is the first to practically enable the verifiability of a two-layer neural network evaluation for image recognition, with less than $1\times$ computation overhead for the client and server, compared to the HE baseline. It also enables the verifiability of a disease prediction result on genomic data with less than $3\times$. In both cases, the communication overhead is at most $2\times$.

In summary, our contributions are the following:
\begin{itemize}
    \item The design of two error-detecting HE encodings that support all the operations enabled by efficient HE schemes over integer plaintexts. We provide an in-depth study of their respective utility tradeoffs.
    \item Several optimizations that reduce the communication and computation overhead induced by our encodings. We present, in particular, a new communication-efficient \textit{polynomial compression protocol} and a client-aided \textit{re-quadratization} technique that reduces both the server's computation overhead and the overall communication overhead.
    \item The implementation of \namecomma, a library that clients can use to detect with high probability a malicious server attacking the HE pipeline. \name is the first open-source and ready-to-use library that enables off-the-shelf secure homomorphic analytics under malicious but rational adversaries. We evaluate \namecomma' performance on five use-cases and show that it outperforms the state of the art~\cite{fiore2014efficiently,ganesh2021rinocchio}. 
\end{itemize}

\section{Problem Statement and Solution Overview}\label{sec:pb}
In this section, we present our system and threat models and the desired objectives, before presenting an overview of our solution.

\descr{System \& Threat Model.} We consider an HE-based computation scenario where one or more clients desire to perform computations on their sensitive data, which they encrypt with an HE scheme. 
The clients employ a computing server that is responsible for carrying out the computations; these range from the entirety of the computation in the case of a single-client~\cite{chen2018logistic,cheon2017privacy,kim2015private}, to aiding in the homomorphic evaluation for multiparty scenarios~\cite{chen2017batched,chen2019efficient,mouchet2021multiparty,chen2017batched,chen2019efficient}. We consider a malicious-but-rational server that tampers with the computations and does not want to be detected (\ie detected only with a negligible probability~\cite{aumann2007security}). 
Our model does not consider that the adversary has access to a decryption oracle.\footnote{Such models, called $\text{IND-CPA}^{\text{D}}$ require additional countermeasures to prevent vulnerabilities (see~\cite{li2021security,cheon2024attacks}).} 
Similarly, our model considers that the adversary cannot perform verification queries and is oblivious to the client's behavior. 
We assume that the client(s) and the server are authenticated to each other. We do not consider network faults in the communication.

\begin{figure}[t]
    \centering
    \resizebox{0.6\columnwidth}{!}{
    \tikzset{every picture/.style={line width=0.75pt}} 

\begin{tikzpicture}[x=0.75pt,y=0.75pt,yscale=-1,xscale=1]

\draw    (211,264.95) .. controls (205.3,276.35) and (200.5,280.58) .. (200.04,298.12) ;
\draw [shift={(200,301)}, rotate = 270] [fill={rgb, 255:red, 0; green, 0; blue, 0 }  ][line width=0.08]  [draw opacity=0] (7.14,-3.43) -- (0,0) -- (7.14,3.43) -- cycle    ;
\draw   (89,246) .. controls (89,245.45) and (89.45,245) .. (90,245) -- (178,245) .. controls (178.55,245) and (179,245.45) .. (179,246) -- (179,270) .. controls (179,270.55) and (178.55,271) .. (178,271) -- (90,271) .. controls (89.45,271) and (89,270.55) .. (89,270) -- cycle ;
\draw  [dash pattern={on 0.84pt off 2.51pt}]  (62,279.75) -- (470,280) ;
\draw   (90,301) .. controls (90,300.45) and (90.45,300) .. (91,300) -- (149,300) .. controls (149.55,300) and (150,300.45) .. (150,301) -- (150,325) .. controls (150,325.55) and (149.55,326) .. (149,326) -- (91,326) .. controls (90.45,326) and (90,325.55) .. (90,325) -- cycle ;
\draw   (160,301) .. controls (160,300.45) and (160.45,300) .. (161,300) -- (219,300) .. controls (219.55,300) and (220,300.45) .. (220,301) -- (220,325) .. controls (220,325.55) and (219.55,326) .. (219,326) -- (161,326) .. controls (160.45,326) and (160,325.55) .. (160,325) -- cycle ;
\draw   (243,301) .. controls (243,300.45) and (243.45,300) .. (244,300) -- (302,300) .. controls (302.55,300) and (303,300.45) .. (303,301) -- (303,325) .. controls (303,325.55) and (302.55,326) .. (302,326) -- (244,326) .. controls (243.45,326) and (243,325.55) .. (243,325) -- cycle ;
\draw    (220,313) -- (237,313) ;
\draw [shift={(240,313)}, rotate = 180] [fill={rgb, 255:red, 0; green, 0; blue, 0 }  ][line width=0.08]  [draw opacity=0] (7.14,-3.43) -- (0,0) -- (7.14,3.43) -- cycle    ;
\draw    (150,313) -- (157,313) ;
\draw [shift={(160,313)}, rotate = 180] [fill={rgb, 255:red, 0; green, 0; blue, 0 }  ][line width=0.08]  [draw opacity=0] (7.14,-3.43) -- (0,0) -- (7.14,3.43) -- cycle    ;
\draw   (79,294.76) -- (230,294.76) -- (230,330.55) -- (79,330.55) -- cycle ;
\draw  [fill={rgb, 255:red, 155; green, 155; blue, 155 }  ,fill opacity=0.43 ] (240,297.76) -- (306,297.76) -- (306,328.24) -- (240,328.24) -- cycle ;
\draw    (70,313) -- (87,313) ;
\draw [shift={(90,313)}, rotate = 180] [fill={rgb, 255:red, 0; green, 0; blue, 0 }  ][line width=0.08]  [draw opacity=0] (7.14,-3.43) -- (0,0) -- (7.14,3.43) -- cycle    ;
\draw   (323,301) .. controls (323,300.45) and (323.45,300) .. (324,300) -- (382,300) .. controls (382.55,300) and (383,300.45) .. (383,301) -- (383,325) .. controls (383,325.55) and (382.55,326) .. (382,326) -- (324,326) .. controls (323.45,326) and (323,325.55) .. (323,325) -- cycle ;
\draw   (393,301) .. controls (393,300.45) and (393.45,300) .. (394,300) -- (452,300) .. controls (452.55,300) and (453,300.45) .. (453,301) -- (453,325) .. controls (453,325.55) and (452.55,326) .. (452,326) -- (394,326) .. controls (393.45,326) and (393,325.55) .. (393,325) -- cycle ;
\draw    (453,313) -- (470,313) ;
\draw [shift={(473,313)}, rotate = 180] [fill={rgb, 255:red, 0; green, 0; blue, 0 }  ][line width=0.08]  [draw opacity=0] (7.14,-3.43) -- (0,0) -- (7.14,3.43) -- cycle    ;
\draw    (383,313) -- (390,313) ;
\draw [shift={(393,313)}, rotate = 180] [fill={rgb, 255:red, 0; green, 0; blue, 0 }  ][line width=0.08]  [draw opacity=0] (7.14,-3.43) -- (0,0) -- (7.14,3.43) -- cycle    ;
\draw    (303,313) -- (320,313) ;
\draw [shift={(323,313)}, rotate = 180] [fill={rgb, 255:red, 0; green, 0; blue, 0 }  ][line width=0.08]  [draw opacity=0] (7.14,-3.43) -- (0,0) -- (7.14,3.43) -- cycle    ;
\draw   (317.5,295.1) -- (459,295.1) -- (459,330.9) -- (317.5,330.9) -- cycle ;
\draw    (211,264.95) .. controls (220.8,281.61) and (332.4,266.62) .. (357.57,297.08) ;
\draw [shift={(359,299)}, rotate = 236.31] [fill={rgb, 255:red, 0; green, 0; blue, 0 }  ][line width=0.08]  [draw opacity=0] (7.14,-3.43) -- (0,0) -- (7.14,3.43) -- cycle    ;
\draw [color={rgb, 255:red, 0; green, 0; blue, 0 }  ,draw opacity=1 ]   (179,258) -- (196,258) ;
\draw [shift={(199,258)}, rotate = 180] [fill={rgb, 255:red, 0; green, 0; blue, 0 }  ,fill opacity=1 ][line width=0.08]  [draw opacity=0] (7.14,-3.43) -- (0,0) -- (7.14,3.43) -- cycle    ;
\draw    (238,264.95) .. controls (246.64,277.43) and (257.12,277.93) .. (268.56,295.69) ;
\draw [shift={(270,298)}, rotate = 239.1] [fill={rgb, 255:red, 0; green, 0; blue, 0 }  ][line width=0.08]  [draw opacity=0] (7.14,-3.43) -- (0,0) -- (7.14,3.43) -- cycle    ;

\draw (421,262) node [anchor=north west][inner sep=0.75pt]  [font=\small] [align=left] {\textit{Offline}};
\draw (421,283) node [anchor=north west][inner sep=0.75pt]  [font=\small] [align=left] {\textit{Online}};
\draw (68,313) node [anchor=east] [inner sep=0.75pt]    {$\mathbf{m}$};
\draw (134,258) node   [align=left] {HE KeyGen};
\draw (120,313) node   [align=left] {Encoder};
\draw (190,313) node   [align=left] {Encrypt};
\draw (273,313) node   [align=left] {HE Eval};
\draw (475,313) node [anchor=west] [inner sep=0.75pt]   [align=left] {$f(\mathbf{m} )$};
\draw (353,313) node   [align=left] {Decrypt};
\draw (423,313) node   [align=left] {Decoder};
\draw (201,258) node [anchor=west] [inner sep=0.75pt]  [color={rgb, 255:red, 0; green, 0; blue, 0 }  ,opacity=1 ]  {$\mathbf{sk}_{\text{HE}} ,\mathbf{evk}$};

\end{tikzpicture}
    }
    \caption{HE pipeline. It returns the homomorphic evaluation of a function $f(\cdot)$ over the encryption of a message $\textbf{m}$.}
    \label{fig:fhepure}
\end{figure}
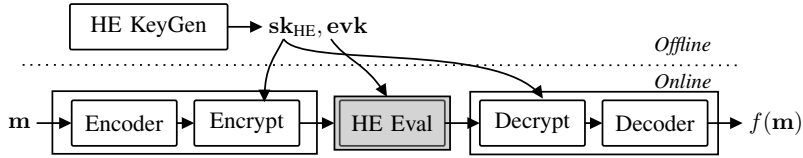

\descr{Objectives.} The client wants (i)~to guarantee the privacy of its input data and the computation output vis-à-vis the server. In particular, the server should learn nothing about the input data and the computation result. The client also wants (ii)~to ensure the correctness of the result with respect to the agreed-upon computation and inputs: the client must be able to detect a cheating server with probability at least $1-2^{-\lambda}$ for a security parameter $\lambda$. 

\begin{figure}[t]
    \centering
    \resizebox{0.6\columnwidth}{!}{
    \input{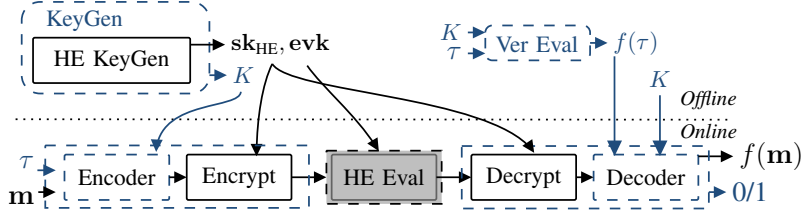}
    }
    \caption{Enhanced HE pipeline. The square boxes correspond to the original HE pipeline (see \S\ref{sec:prelim:fhe} and Fig.~\ref{fig:fhepure}) and the dotted boxes are the new components offering verification capabilities. The grey box represents the computing server. For a message $\textbf{m}$ associated with a label $\tau$ and for a function $f(\cdot)$, the verification pipeline checks if the Eval. step performed by the computing server was executed correctly. $\mathbf{sk}_{\text{HE}}$ and $\mathbf{evk}$ are the HE secret key and evaluation key respectively. $K$ is the encoder's secret key.}
    \label{fig:fhe}
\end{figure}

\descr{Solution Overview.} To protect data privacy during outsourcing and the subsequent server computation (\ie Objective (i)), the client uses state-of-the-art lattice-based HE to encrypt it (\S\ref{sec:prelim:fhe}). This protects also the privacy of the client, in the absence of a decryption and verification oracle (\ie the server does not obtain any information after a decryption). To ensure the correctness of the server's computation on the outsourced data (\ie Objective (ii)), we embed the verification of the computations into the plaintext space. During outsourcing, the client encodes its data using an error-detection encoder, encrypts and sends it (see Figure~\ref{fig:fhe}). We design two encoders with error-detecting capabilities that respect the homomorphic operations in the plaintext space. Combined with the HE scheme, they emulate homomorphic authenticators (\S\ref{sec:prelim:vche}) that enable client-based verification capabilities and act in lieu of the classic HE pipeline (by adapting the HE evaluation and decryption to the new plaintext encoders -- see Figure~\ref{fig:fhepure} vs. Figure~\ref{fig:fhe}). After the server computation, the client decrypts and verifies the integrity of the encodings, thus vouching for the computation correctness with high probability. The first encoder, which is based on replication, mixes replicas of the data with challenge values in an extended vector (see Figure~\ref{fig:vche1}) that is encrypted by the client. After the server computation, the client can check its correctness by decrypting and inspecting the resulting extended vector. This encoder is detailed in \S\ref{sec:vche1}. The second one, which is based on a polynomial encoding, encodes the message as a bivariate polynomial (see Figure~\ref{fig:vche2}) that is encrypted by the client. After the server computation, the client verifies its correctness by decrypting and evaluating the resulting polynomial on a secret point. This is presented in \S\ref{sec:vche2}. The two encodings support any HE operation and prevent malicious servers from tampering with the requested computations undetected while achieving different efficiency trade-offs that we analyze in \S\ref{sec:eval}.

\descr{Verification Outcome.}
With our solution, a client is able to detect with high probability if the server misbehaves. Thus, malicious but rational servers are strongly discouraged from tampering with the requested computations. Yet, similarly to prior HE works, our solution would be vulnerable if the adversary learns the outcome of the decryptions~\cite{cheon2024attacks, li2021security} and the result of the verification. Thus, our threat model does not consider a decryption oracle and the client's behavior following a verification should be independent from the verification result. To ensure this there is a need for orthogonal countermeasures at the application layer on top of our solution: \eg single query-system, dummy queries, etc. In the remainder of the paper, and w.l.o.g., we assume the client sends a constant number of independent computation requests at regular intervals (either legitimate or dummy ones) to the server and then resets the system. This way, the server is oblivious to the verification result (see Appendix~\ref{ap:disc:outcome} for a formal description).

\section{Preliminaries}\label{sec:prelim}
We introduce some key components used in this work. Let $a \leftarrow \chi$ denote that $a$ is sampled uniformly at random from a distribution $\chi$; a vector (resp. polynomial) be denoted by a boldface letter, \eg $\mathbf{x}$, with $\mathbf{x}[i]$ its $i$-th element (resp. coefficient). Let $[n]$ refer to the set of integers $\{k; 0 {<} k{\leqslant} n\}$ and $[0{:}n]$ the set $\{0\}{\cup} [n]$. The security parameter is an integer denoted by $\lambda$. For a prime $t$, denote $\mathbb{Z}_t$ its finite field and $\mathbb{Z}^*_t$ its non-zero elements. Let $[m]_q$ denote $m\!\!\mod\!q$.

\subsection{Homomorphic Encryption}\label{sec:prelim:fhe}
Homomorphic encryption (HE) enables the execution, without decryption, of certain operations on ciphertexts. When decrypted, the resulting ciphertext reveals the result of the computation as if it were executed on the plaintexts. Trending lattice-based HE schemes, \eg BFV~\cite{fan2012somewhat,halevi2019improved} and BGV~\cite{brakerski2014leveled,kim2021revisiting}, rely on the hardness of the learning with errors (LWE)~\cite{regev2009lattices} or ring-LWE (RLWE)~\cite{lyubashevsky2010ideal} problems. Such schemes approach fully homomorphic encryption (FHE) as they can support an unbounded number of additions and multiplications due to a bootstrapping operation~\cite{kim2021general}.

Our work focuses on lattice-based HE schemes that support modular arithmetic in a field $\mathbb{Z}_t$, \eg the BGV and BFV schemes that are under standardization~\cite{HomomorphicEncryptionSecurityStandard}. These schemes share the same plaintext algebra and differ only in the plaintext data-representation and encryption-noise management (see~\cite{kim2021revisiting}). For completeness, we describe briefly the BFV scheme used in this work.  
The plaintext space is $\mathcal{R}_t{=}\mathbb{Z}_t[X]/\!( X^N\!{+}1)$, with $t$ a prime and $N$ a power-of-two polynomial degree. Analogously, the cipherspace is $\mathcal{R}_q^2$ with $q\gg t$. We denote by $\chi_k$ and $\chi_e$ the key and error distribution in $\mathcal{R}_q$.

\descr{HE Encoders.} As shown in Figure~\ref{fig:fhepure}, a plaintext encoder converts a vector of scalar values into a polynomial element in $\mathcal{R}_t$. Several encoders exist: \eg scalar, integer, and fractional ones (see \cite[\S7]{laine2017simple}). In this work, we employ an encoder called \textit{batching} that enables Single Instruction, Multiple Data (SIMD) operations. It converts a vector in $\mathbb{Z}_t^N$ to a polynomial in $\mathcal{R}_t$. To enable server-computation verifiability (see \S\ref{sec:pb}), we extend this encoder to introduce error-detection capabilities (see \S\ref{sec:vche1}, \S\ref{sec:vche2}). We term \textit{slot} a component of the plaintext vector before its encoding as a polynomial. 

The main BFV algorithms are:

\descr{BFV.KeyGen}($1^\lambda$)$\rightarrow \mathbf{sk}_{\text{HE}}, \mathbf{pk}_{\text{HE}}, \mathbf{evk}_{\text{HE}}$: Choose a low-norm secret key $\mathbf{s} \!\leftarrow\! \chi_k$ and set $\mathbf{sk}_{\text{HE}} {:=} (1,\mathbf{s})$. The public key is defined as $\mathbf{pk}_{\text{HE}}\!:=\!(\mathbf{b},\mathbf{a})$ where $\mathbf{a}$ is sampled uniformly at random from $\mathcal{R}_q$ and $\mathbf{b}{:=}[-(\mathbf{a}\cdot\mathbf{s}+\mathbf{e})]_q{\in} \mathcal{R}_q$ with $\mathbf{e}\!\leftarrow\!\chi_e$. For the evaluation, define an evaluation key $\mathbf{evk}_{\text{HE}}$ that comprises the relinearization and the rotation keys. 
Output $\mathbf{sk}_{\text{HE}}$, $\mathbf{pk}_{\text{HE}}$, and $\mathbf{evk}_{\text{HE}}$.

\descr{BFV.Enc}($\mathbf{p};\mathbf{pk}_{\text{HE}}$)$\rightarrow \mathbf{c}$: For a message $\mathbf{m} \in \mathbb{Z}_t^N$ encoded as a polynomial $\mathbf{p}\in \mathcal{R}_t$, sample $\mathbf{u} \leftarrow \chi_k$ and $\mathbf{e}_0', \mathbf{e}_1' \leftarrow \chi_e$. Output $\mathbf{c}{:=}[\mathbf{u} \cdot \mathbf{pk}_{\text{HE}} {+} (\mathbf{e}_0'+\lceil q/t \rfloor \cdot \mathbf{p}, \mathbf{e}_1')]_q$. Thus, a fresh ciphertext comprises two polynomials in $\mathcal{R}_q$, \ie $\mathbf{c}{=}(\mathbf{c}_0,\mathbf{c}_1)$.

\descr{BFV.Dec}($\mathbf{c};\mathbf{sk}_{\text{HE}}$)$\rightarrow \!\mathbf{m}$: Decode $[\lceil t/q\!\cdot\! [\!\langle \mathbf{sk}_{\text{HE}}, \mathbf{c}\rangle\!]_q \rfloor]_t$ to $\mathbf{m}$.

BFV is a \textit{canonical} HE scheme \ie it supports circuit evaluation and composition. Let us now present the supported operations.

\descr{BFV.Add}($\mathbf{c},\! \hat{\mathbf{c}}$): A homomorphic addition simply adds the ciphertext vectors in $\mathcal{R}_q^2$. Given $\mathbf{c}$ and $\hat{\mathbf{c}}$, it outputs $[\mathbf{c}{+} \hat{\mathbf{c}}]_q$.

\descr{BFV.Mul}($\mathbf{c},\hat{\mathbf{c}}; \mathbf{evk}_{\text{HE}}$): The multiplication requires a relinearization procedure. Given two ciphertexts $\mathbf{c}{=}(\mathbf{c}_0,\mathbf{c}_1)$ and $\hat{\mathbf{c}}{=}(\hat{\mathbf{c}}_0,\hat{\mathbf{c}}_1)$, it does the following:
\begin{enumerate}[leftmargin=*]
    \item \textbf{Tensoring:} Computes $\mathbf{c}_{0}' {:=} \mathbf{c}_0 \hat{\mathbf{c}}_0$, $\mathbf{c}_{1}'{:=}\mathbf{c}_0\hat{\mathbf{c}}_1 {+} \mathbf{c}_1 \hat{\mathbf{c}}_0$, and $\mathbf{c}_2':=\mathbf{c}_1\hat{\mathbf{c}}_1 \in \mathcal{R}$ \emph{without modular reduction}. Note that this represents the convolution between the two input ciphertexts.
    \item \textbf{Relinearization:} Sets $\mathbf{c}_i^*{:=}[\lceil t/q \cdot \mathbf{c}_i' \rfloor]_q$, for $i\in\{0,1,2\}$. Then it uses the relinearization key to convert the ciphertext $\mathbf{c}^*$ of three polynomials to a ciphertext comprising only two polynomials. This operation is not arithmetic and cannot be performed over $\mathcal{R}_q$ (see \cite{halevi2019improved}). 
\end{enumerate}\vspace{0.2em}

\noindent To fully support SIMD operations, a homomorphic rotation operation (\ie $\textbf{BFV.Rot}_r()$) rotates the plaintext vector slots by using rotation keys that are created and shared. Each rotation key is associated with a specific rotation step (\ie $r$) and stored in the evaluation key $\mathbf{evk}_{\text{HE}}$. Additional operations, \eg bootstrapping or key-switching, provide further functionalities without affecting the plaintext. All these operations cannot be represented by an arithmetic circuit in $\mathcal{R}_q$ as they require arithmetic over an extended ring or require rounding operations. In the following, we denote by $\textbf{BFV.Eval}(f(\cdot), (\mathbf{c}_1, \!..., \mathbf{c}_n);\mathbf{evk}_{\text{HE}})$ the SIMD evaluation of the plaintext function $f(\cdot)$ on the ciphertexts $(\mathbf{c}_1, \!..., \mathbf{c}_n)$. 

\subsection{Verifiable Programs}\label{sec:prelim:verprog} 
To enable computation verification, it is necessary to clearly identify the inputs, the output, and the function being evaluated. We briefly recall here a formalization of \textit{identifiers} and \textit{programs} for verifiable computation~\cite{backes2013verifiable,gennaro2013fully}.
Any possible input message $m$ is represented by an identifier $\tau$. This identifier can be seen as a string uniquely identifying the data in a database. 
A \textit{program} corresponds to the application of a function (\ie circuit) $f(\cdot)$ to inputs associated with specific identifiers: it is labeled as a tuple $\mathcal{P}{=}(f(\cdot), (\tau_1, \!... , \tau_n))$. Correctness of a program ensures that, given the identified inputs, the output of the function is the expected one without corruption. A program is said to be \textit{well-defined} if all inputs contributing to the computation are authenticated (see Catalano and Fiore for the formal definition~\cite{catalano2013practical}).

\subsection{Homomorphic Authenticators}\label{sec:prelim:vche} 
Homomorphic authenticators (HA)~\cite{gennaro2013fully} are cryptographic schemes used to verify computation integrity. They consist of four probabilistic polynomial time (PPT) procedures: the key generation (\textbf{HA.KeyGen}), the authentication (\textbf{HA.Auth}) of the input data, the evaluation (\textbf{HA.Eval}) of an agreed-upon function on authenticated data, and the verification (\textbf{HA.Ver}) of the claimed output. Correctly instantiated homomorphic authenticators provide authentication and evaluation correctness as well as security (see Appendix~\ref{ap:HA} for a formal description). In the following sections, we show how to instantiate an HA using a verifiable HE-encoding. As mentioned in \S\ref{sec:pb}, by adapting the HE pipeline to work with our new plaintext encoders, we emulate the HA procedures for the authentication, the evaluation, and the verification. Figure~\ref{fig:fhe} displays the modifications performed to the original HE pipeline as dashed boxes.

\section{Replication Encoding}\label{sec:vche1}
We first design a replication-based verifiable encoding (\S\ref{sec:vche1:encoder}) that takes advantage of the native HE batching to introduce error-detection and redundancy in the data. Then, we construct an authenticator (\S\ref{sec:vche1:def}) following the footprint of homomorphic MACs~\cite{gennaro2013fully,chung2010improved}.

\begin{figure}[t]
    \centering
    \resizebox{0.6\columnwidth}{!}{
    \tikzset{every picture/.style={line width=0.75pt}} 

\begin{tikzpicture}[x=0.75pt,y=0.75pt,yscale=-1,xscale=1]

\draw   (80,102) .. controls (80,95.37) and (85.37,90) .. (92,90) -- (368,90) .. controls (374.63,90) and (380,95.37) .. (380,102) -- (380,138) .. controls (380,144.63) and (374.63,150) .. (368,150) -- (92,150) .. controls (85.37,150) and (80,144.63) .. (80,138) -- cycle ;
\draw   (156.6,114) .. controls (156.6,111.79) and (158.39,110) .. (160.6,110) -- (314.6,110) .. controls (316.81,110) and (318.6,111.79) .. (318.6,114) -- (318.6,126) .. controls (318.6,128.21) and (316.81,130) .. (314.6,130) -- (160.6,130) .. controls (158.39,130) and (156.6,128.21) .. (156.6,126) -- cycle ;
\draw    (325,120) -- (350,120) ;
\draw [shift={(350,120)}, rotate = 180] [fill={rgb, 255:red, 0; green, 0; blue, 0 }  ][line width=0.08]  [draw opacity=0] (7.14,-3.43) -- (0,0) -- (7.14,3.43) -- cycle    ;
\draw [color={rgb, 255:red, 155; green, 155; blue, 155 }  ,draw opacity=1 ]   (207.99,136.5) -- (207.99,123.5) ;
\draw [shift={(207.99,120.5)}, rotate = 90] [fill={rgb, 255:red, 155; green, 155; blue, 155 }  ,fill opacity=1 ][line width=0.08]  [draw opacity=0] (7.14,-3.43) -- (0,0) -- (7.14,3.43) -- cycle    ;
\draw [color={rgb, 255:red, 155; green, 155; blue, 155 }  ,draw opacity=1 ]   (312.99,136.53) -- (312.99,123.53) ;
\draw [shift={(312.99,120.53)}, rotate = 90] [fill={rgb, 255:red, 155; green, 155; blue, 155 }  ,fill opacity=1 ][line width=0.08]  [draw opacity=0] (7.14,-3.43) -- (0,0) -- (7.14,3.43) -- cycle    ;
\draw [color={rgb, 255:red, 155; green, 155; blue, 155 }  ,draw opacity=1 ]   (188,136.5) -- (188,123.5) ;
\draw [shift={(188,120.5)}, rotate = 90] [fill={rgb, 255:red, 155; green, 155; blue, 155 }  ,fill opacity=1 ][line width=0.08]  [draw opacity=0] (7.14,-3.43) -- (0,0) -- (7.14,3.43) -- cycle    ;
\draw    (170,100) -- (170,107) ;
\draw [shift={(170,110)}, rotate = 270] [fill={rgb, 255:red, 0; green, 0; blue, 0 }  ][line width=0.08]  [draw opacity=0] (5.36,-2.57) -- (0,0) -- (5.36,2.57) -- cycle    ;
\draw    (230,100) -- (230,107) ;
\draw [shift={(230,110)}, rotate = 270] [fill={rgb, 255:red, 0; green, 0; blue, 0 }  ][line width=0.08]  [draw opacity=0] (5.36,-2.57) -- (0,0) -- (5.36,2.57) -- cycle    ;
\draw    (290,100) -- (290,107) ;
\draw [shift={(290,110)}, rotate = 270] [fill={rgb, 255:red, 0; green, 0; blue, 0 }  ][line width=0.08]  [draw opacity=0] (5.36,-2.57) -- (0,0) -- (5.36,2.57) -- cycle    ;
\draw    (270,100) -- (270,107) ;
\draw [shift={(270,110)}, rotate = 270] [fill={rgb, 255:red, 0; green, 0; blue, 0 }  ][line width=0.08]  [draw opacity=0] (5.36,-2.57) -- (0,0) -- (5.36,2.57) -- cycle    ;
\draw    (170,100) -- (324,100) ;
\draw [color={rgb, 255:red, 155; green, 155; blue, 155 }  ,draw opacity=1 ]   (253.05,136.45) -- (253.05,123.45) ;
\draw [shift={(253.05,120.45)}, rotate = 90] [fill={rgb, 255:red, 155; green, 155; blue, 155 }  ,fill opacity=1 ][line width=0.08]  [draw opacity=0] (7.14,-3.43) -- (0,0) -- (7.14,3.43) -- cycle    ;
\draw    (104,120) -- (120,120) ;
\draw [shift={(120,120)}, rotate = 180] [fill={rgb, 255:red, 0; green, 0; blue, 0 }  ][line width=0.08]  [draw opacity=0] (7.14,-3.43) -- (0,0) -- (7.14,3.43) -- cycle    ;

\draw (93.5,120) node    {$\mathsf{m}, \tau\;$};
\draw (139,120) node    {$\mathbf{M} =$};
\draw (168.5,120) node    {$\mathsf{m}$};
\draw (228.5,120) node    {$\mathsf{m}$};
\draw (268.74,120.2) node    {$\mathsf{m}$};
\draw (288.74,120.2) node    {$\mathsf{m}$};
\draw (187.5,120) node    {$\cdot $};
\draw (207.52,120.2) node    {$\cdot $};
\draw (252.5,120) node    {$\cdot $};
\draw (312.52,120.2) node    {$\cdot $};
\draw (357.35,120.2) node    {$\mathbf{p}$};
\draw (361,131) node    {$\in \mathcal{R}_{t}$};
\draw (171.5,142) node  [font=\scriptsize]  {$F_{K} (\tau ,2)$};
\draw (211.5,142) node  [font=\scriptsize]  {$F_{K} (\tau ,3)$};
\draw (258.5,142) node  [font=\scriptsize]  {$F_{K} (\tau ,5)$};
\draw (308.5,142) node  [font=\scriptsize]  {$F_{K} (\tau ,8)$};
\draw (343.5,100) node    {$\notin S$};

\end{tikzpicture}
    }
    \caption{{Replication Encoding. A message $\mathsf{m}$ with identifier $\tau$ is encoded as a vector $\mathbf{M}$ with challenge values (using the PRF $F_K(\cdot)$) for indices in the challenge set $S$ and replications of $\mathsf{m}$ for all the others. For ease of presentation, here $\mathsf{m}{\in}\mathbb{Z}_t$ and $\lambda {=} 8$.}}
    \label{fig:vche1}
\end{figure}
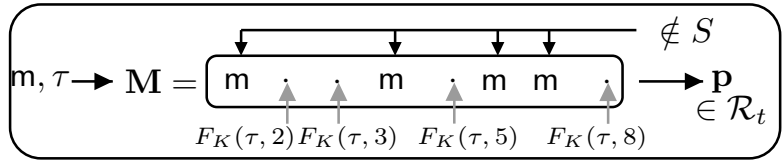

\subsection{Replication-Based Encoding}\label{sec:vche1:encoder}

On input a vector $\mathbf{m}$ with identifier $\boldsymbol{\tau}$, a power-of-two security parameter $\lambda$, a challenge set $S\subset [\lambda]$ of size $\lambda/2$, and a pseudorandom function (PRF) $F_K(\cdot)$, the replication-based encoder returns an encoding as follows: For each scalar value of $\mathbf{m}$ (say the $i$-th), an extended vector $\mathbf{M}_i$ of length $\lambda$ is created. For all indices of $\mathbf{M}_i$ in $S$, the vector is filled with a challenge value obtained from the PRF and the identifier $\boldsymbol{\tau}[i]$ associated with $\mathbf{m}[i]$. The remaining empty elements of $\mathbf{M}_i$ are filled with replicas of the initial message $\mathbf{m}[i]$. All the extended vectors are then concatenated to create the encoding $\mathbf{M}$. The encoding is then parsed as a concatenation of HE plaintexts and encoded in $\mathcal{R}_t$. A simplified version of the replication encoding for one scalar value is presented in Figure~\ref{fig:vche1}, and the encoder details are shown in Scheme~\ref{ecd:REP}.

\begin{figure}[t]
\refstepcounter{scheme}\label{ecd:REP}
\scheme{\textsc{\footnotesize{Replication-Based Encoder in}} \footnotesize{$\mathcal{R}_t{=}\mathbb{Z}_t[X]/( X^N\!\!{+}1)$}}{
~
$\textbf{ReplicationEncoder}(\mathbf{m}, \boldsymbol{\tau}; \lambda, S, F_K(\cdot))$:
\begin{enumerate}[leftmargin=*, topsep=0pt, itemsep=0ex]
    \item Set an empty vector $\mathbf{M}{=}()$.
    \item For the message $\mathbf{m}{\in}\mathbb{Z}_t^N$ with identifiers in $\boldsymbol{\tau}$. $\forall i{\in}[N]$:
    \begin{itemize}[leftmargin=0.3pt, topsep=0pt, itemsep=0ex]
        \item Set a vector $\mathbf{M}_i \in  \mathbb{Z}_t^\lambda$  s.t. $\mathbf{M}_i[j] =  \mathbf{m}[i] $ if $j \notin S$ and  $\mathbf{M}_i[j] {=}F_K(\boldsymbol{\tau}[i],j)$ otherwise.
        \item Set the concatenations $\mathbf{M}{=}\mathbf{M}|\mathbf{M}_i$.
    \end{itemize}
    \item Process $\mathbf{M}{\in} \mathbb{Z}_t^{\lambda N}\!$\!\!\, as $(\!\mathbf{p}_1\!|...|\mathbf{p}_{\lambda}\!) {\in} \mathcal{R}_t^{\lambda}$ using the \textit{batching encoder}.
    \item Return $(\mathbf{p}_1|\dots|\mathbf{p}_{\lambda})$.
\end{enumerate}
}
\end{figure}

\subsection{Replication-Based Authenticator}\label{sec:vche1:def}
We combine the replication-based encoding with an HE scheme to obtain a homomorphic authenticator~\cite{gennaro2013fully}. As we change the plaintext encoder, the evaluation and decryption algorithms of the HE pipeline need to be adapted. We present how in the following.

Let $F_K{:} \mathcal{I}{\rightarrow}\mathbb{Z}_t$ be a variable length pseudorandom function (PRF) and $\text{HE}$ a secure homomorphic encryption scheme as in \S\ref{sec:prelim:fhe} with plaintext space $\mathcal{R}_t{=}\mathbb{Z}_t[X]/( X^N\!{+}1)$ of degree $N$. Let $\mathcal{F}$ be a set of admissible polynomial functions over $\mathbb{Z}_t$ with degree $\mathrm{d}\ll t$ and a non-trivial and varied image space (\ie for which the image cannot be trivially guessed w.p. more than 1/3). 
We define a \textit{Replication Encoding}-based authenticator (REP) as in Scheme~\ref{scheme:REP} that we describe here.

\descr{$\text{REP.KeyGen}(1^\lambda)$}$\rightarrow \!(\mathbf{evk}, \mathbf{sk})$: For a power-of-two security parameter $\lambda$, this procedure sets up the PRF with key $K$ and instantiates the $\text{HE}$ scheme such that both achieve at least $\lambda$-bits security. It also generates the encryption and evaluation keys. The latter comprise the relinearization key and the rotation keys. It picks at random a challenge set of indices $S\subset [\lambda]$ subject to $|S|{=}\frac{\lambda}{2}$. Finally, it sets the authenticator evaluation and secret keys.

\descr{$\text{REP.Auth}(\mathbf{m}, \boldsymbol{\tau}; \mathbf{sk})$}$\rightarrow \!\boldsymbol{\sigma}$: For an input vector $\mathbf{m} {\in} \mathbb{Z}_t^N$ with identifier $\boldsymbol{\tau}$ it proceeds sequentially. It first encodes $\mathbf{m}$ by calling the encoding procedure $\textbf{ReplicationEncoder}(\mathbf{m}, \boldsymbol{\tau}; \lambda, S, F_K(\cdot))$ which returns a list of plaintexts over $\mathcal{R}_t$ (\underline{Encode}). Then, it encrypts each encoded plaintext and appends it to a list of ciphertexts $\mathbf{c}$ (\underline{\smash{Encrypt}}). Finally, it outputs the authentication $\boldsymbol{\sigma}{:=}\mathbf{c}$.

\descr{$\text{REP.Eval}(f(\cdot), \vv{\boldsymbol{\sigma}}; \mathbf{evk})$}$\rightarrow \!\boldsymbol{\sigma}'$: For a polynomial function $f(\cdot){\in}\mathcal{F}$ and $n$ authenticated input vectors represented by $\vv{\boldsymbol{\sigma}}{=}(\boldsymbol{\sigma}_1, \!..., \boldsymbol{\sigma}_n)$, it computes $\mathbf{c}'$ by evaluating homomorphically the corresponding SIMD circuit on the ciphertexts $(\mathbf{c}_1, \!..., \mathbf{c}_n)$. All operations are executed slot-wise, and rotations steps are increased by a factor $\lambda$. It outputs $\boldsymbol{\sigma}' {:=} \mathbf{c}'$. 

\descr{$\text{REP.Ver}(\mathcal{P}, \boldsymbol{\sigma}'; \mathbf{sk})$}$\rightarrow \{0,1\}$: It parses the labeled program $\mathcal{P} {=} (\allowbreak f(\cdot), (\boldsymbol{\tau}_1, \!..., \boldsymbol{\tau}_n))$. 

Offline, (i)~it checks if the function is admissible (\ie $f {\in} \mathcal{F}$) and (ii)~it pre-computes the challenge values by using both the identifier and the PRF, and it evaluates on them the function $f$ in the plaintext space. If the function is not admissible or all the challenge evaluations return the same value, it aborts.  
This prevents the verification of an inadmissible function or of a badly formatted encoding. In particular, the abort mechanism prevents the server from being able to blindly manipulate the extended vector in the event that all the challenge values are the same after the evaluation. To circumvent the second abort, the client requests a new program by adding a new variable: \ie $\mathcal{P}' {=} (f+Id,\allowbreak (\boldsymbol{\tau}_1, \!..., \boldsymbol{\tau}_n), \boldsymbol{\tau}_y))$ that returns $f(x_1, \dots, x_n)+y$ with $y=0$. This ensures w.h.p. (assuming a large enough $t$) that the new challenges do not evaluate to the same value. 

Online, it decrypts the ciphertext $\mathbf{c}'$ in $\boldsymbol{\sigma}'$ and decodes it to the plaintext vector $\mathbf{M}^*$ and checks if for all slots $j {\in} S$ the output matches the pre-computed challenge for this slot. Moreover, it ensures that all the other slots $j {\in} [\lambda]\setminus S$ evaluate to the same value. If the above checks pass, it outputs $1$ (\ie it accepts).

\begin{figure}[!t]
\refstepcounter{scheme}\label{scheme:REP}
\scheme{\textsc{\footnotesize{Replication Encoding-Based Authenticator}}}{\vspace{-0.1cm}
~
    \begin{itemize}[leftmargin=*]
    \itemsep0.7em
        \item $\textbf{REP.KeyGen}(1^\lambda)$: For a power-of-two security parameter $\lambda$. 
        \begin{enumerate}[topsep=0pt, itemsep=0ex]
            \item Choose a PRF key $K \!{\leftarrow} \{0,1\}\!^*$ for at least $\lambda$-bits security. 
            \item Set the HE keys $(\mathbf{pk}_{\text{HE}}, \mathbf{evk}_{\text{HE}}, \mathbf{sk}_{\text{HE}}){=} \text{HE.KeyGen}(1^{\lambda})$ for at least $\lambda$-bits security.
            \item Randomly sample a challenge set $S\subset [\lambda]$ s.t. $|S|{=}\lambda/2$.
            \item Return $\mathbf{evk}\!\!=\!\!(\mathbf{pk}_{\text{HE}},\mathbf{evk}_{\text{HE}})$ and $\mathbf{sk} \!\!=\!\! (K,\!S, \mathbf{sk}_{\text{HE}})$.
        \end{enumerate}
        \item $\textbf{REP.Auth}(\mathbf{m}, \boldsymbol{\tau}; \mathbf{sk})$: For ${\mathbf{m}\in \mathbb{Z}_t^N}$ with identifiers in $\boldsymbol{\tau}$.
        \begin{itemize}[leftmargin=*, topsep=0pt, itemsep=0ex]
            \item \underline{Encode}:$(\mathbf{p}_1|\!\,...\,\!|\mathbf{p}_{\lambda}\!){=}$\textbf{ReplicationEncoder}($\mathbf{m}, \!\boldsymbol{\tau};\lambda, \!S, \!F_K(\cdot)$)
            \item \underline{\smash{Encrypt}}:
            \begin{enumerate}[topsep=0pt, itemsep=0ex]
            \item Set an empty list $\mathbf{c}{=}()$.
            \item $\forall i \in [\lambda]$, set $\mathbf{c}=\mathbf{c}| \text{HE.Enc}(\mathbf{p}_i; \mathbf{pk}_{\text{HE}})$
            \item Return $\boldsymbol{\sigma} {:=} \mathbf{c}$.
        \end{enumerate}
        \end{itemize}
        \item $\textbf{REP.Eval}(f(\cdot), \vv{\boldsymbol{\sigma}}; \mathbf{evk})$: For a function $f(\cdot)$ to be computed over previously authenticated encrypted inputs stored in $\vv{\boldsymbol{\sigma}}$:
        \begin{enumerate}[topsep=0pt, itemsep=0ex]
            \item Parse the authenticated inputs $\vv{\boldsymbol{\sigma}}{=}(\boldsymbol{\sigma}_1, \!..., \boldsymbol{\sigma}_n){=}(\boldsymbol{c}_1, \!..., \boldsymbol{c}_n)$.
            \item Evaluate $\mathbf{c}' = \text{HE.Eval}(f(\cdot), (\mathbf{c}_1, \!..., \mathbf{c}_n); \mathbf{evk}_{\text{HE}})$.
            \item Return $\boldsymbol{\sigma}' {:=} \mathbf{c}'$.
        \end{enumerate}
        \item $\textbf{REP.Ver}(\mathcal{P}, \boldsymbol{\sigma}'; \mathbf{sk})$: Parse the target program $\mathcal{P} {=} (f,\allowbreak (\boldsymbol{\tau}_1, \!..., \boldsymbol{\tau}_n))$ and $\mathbf{c}'=\boldsymbol{\sigma}'$.
        Check the function $f(\cdot)$ and pre-compute the challenge values:
        \begin{enumerate}[topsep=0pt, itemsep=0ex]
            \item If $f{\notin}\mathcal{F}$ then abort.
            \item $\forall i {\in} S, \, \forall j{\in} [n], \, \forall k{\in} [N/\lambda]$, set $r_{i,j,k}{=}F_K(\boldsymbol{\tau}_j[k],i)$.
            \item $\forall i {\in} S$ compute on the challenges $\tilde{r}_{i} = f(\{r_{i,j,k}\}_{j {\in}[n]}^{k{\in}[N/\lambda]})$\newline
            \textit{w.l.o.g. the output is a scalar value stored in the first extended vector.} 
            \item If $\forall i {\in} S$ $\tilde{r}_{i} = \tilde{r}$, then abort.
        \end{enumerate}
        In the online phase:
        \begin{enumerate}[topsep=0pt, itemsep=0ex]
            \item Decode the decryption $\mathbf{p}=\!\! \text{HE.Dec}(\mathbf{c}'\!;\mathbf{sk}_{\text{HE}})$ using the \textit{batching decoder} and interpret the resulting vector as $\mathbf{M}^*\!\!\!=\!\!(\mathbf{M}_1,\allowbreak \!..., \!\mathbf{M}_l)$ with $l{=}N/\lambda$.
            \item $\forall i {\in} S$, check if $\mathbf{M}_1[i]{=}\tilde{r}_{i}$. If not, return $0$.
            \item Check that $\forall i {\in} [\lambda]\setminus S$, all $\mathbf{M}_1[i]$ have the same value. If not, return $0$.
            \item Return $1$ (\ie accept).
        \end{enumerate}
    \end{itemize}
}
\end{figure}

\descr{Correctness.} Authentication correctness follows directly from the correctness of the HE scheme, and evaluation correctness follows from the canonical property and correctness of the HE scheme (\S\ref{sec:prelim:fhe}).  

\descr{Security.} Within the scope of our threat model (\S\ref{sec:pb}), the following theorem states that a misbehaving computing server has only a negligible probability of tampering, without getting caught, with the computation requested by the client. 

\begin{theorem}\label{th:REP}
Let $\lambda$ be a power-of-two security parameter. If the PRF $F_K$ and the canonical HE scheme are at least $\lambda$-bit secure, then for any admissible program $\mathcal{P}$, $\text{REP}$ as in Scheme~\ref{scheme:REP} is a secure authenticator and a PPT adversary has a probability of successfully cheating the verification negligible in $\lambda$. 
\end{theorem}

\descrit{Proof Intuition:} The security follows by the security of the PRF and by the semantic security of the HE scheme: the server cannot distinguish which slots hold a replicated value or what the challenge values are. In fine, the adversary can cheat without being detected with only a negligible probability. The formal proof is presented in Appendix~\ref{ap:vche1}.

\descr{Overhead.} We observe that REP induces, in the worst case, communication and computational overhead that is linear in the security parameter $\lambda$ for both the client and the server. Indeed, a vector of $N$ values needs to be encoded as $\lambda$ separate ciphertexts and $\lambda/2$ challenges need to be pre-computed before the verification. In comparison with the original homomorphic MAC~\cite{gennaro2013fully}, REP does not modify the encryption procedure. Compared to computation delegation~\cite{chung2010improved}, REP enables verification of computations over unknown (but identified by $\boldsymbol{\tau}$) inputs.  

\subsection{Optimizing the Verification}\label{sec:vche1:PRF} 
REP relies on both the generation of challenges and the circuit-evaluation on these values. As these are independent of the inputs (only their identifiers are required), they can be pre-processed and computed offline. The computation on the challenges can be outsourced to any entity that does not collude with the computing server (or once the computing server has committed to the result of the computation). Any standard verifiable computation techniques, such as interactive proofs~\cite{goldwasser2015delegating}, SNARKs~\cite{bunz2018bulletproofs} or
STARKs~\cite{ben2019aurora}, could be employed to ensure the correct computation (see Appendix~\ref{ap:disc:snarks}). 
As these techniques operate on cleartext data, we deem this pre-processing orthogonal to our work. 
Additionally, similarly to prior work on computation verification~\cite{backes2013verifiable,fiore2012publicly}, one can rely on a closed-form PRF~\cite{benabbas2011verifiable} to enable the partial re-use of the challenges for evaluation over different datasets. Depending on the function under evaluation, this could accelerate the verification at increased costs for generating the encoding and the challenges. 
Because this pre-processing is independent of the encodings themselves, we omit it in the experimental evaluation (\S\ref{sec:eval}). 
These observations also hold for the second encoder that we present next.

\section{Polynomial Encoding}\label{sec:vche2} 
We design a more compact encoding (\S\ref{sec:vche2:encoder}) and construct an authenticator that requires fewer challenge values to be verified (\S\ref{sec:vche2:def}) following Catalano and Fiore's information-theoretic MAC~\cite{catalano2013practical}.

\subsection{Polynomial-Based Encoding}\label{sec:vche2:encoder}

\begin{figure}[!t]
    \centering
    \resizebox{0.6\columnwidth}{!}{
    \tikzset{every picture/.style={line width=0.75pt}} 

\begin{tikzpicture}[x=0.75pt,y=0.75pt,yscale=-1,xscale=1]

\draw    (150,140) -- (197,140) ;
\draw [shift={(200,140)}, rotate = 180] [fill={rgb, 255:red, 0; green, 0; blue, 0 }  ][line width=0.08]  [draw opacity=0] (7.14,-3.43) -- (0,0) -- (7.14,3.43) -- cycle    ;
\draw    (130,150) -- (130,157) ;
\draw [shift={(130,160)}, rotate = 270] [fill={rgb, 255:red, 0; green, 0; blue, 0 }  ][line width=0.08]  [draw opacity=0] (5.36,-2.57) -- (0,0) -- (5.36,2.57) -- cycle    ;
\draw    (220,150) -- (220,157) ;
\draw [shift={(220,160)}, rotate = 270] [fill={rgb, 255:red, 0; green, 0; blue, 0 }  ][line width=0.08]  [draw opacity=0] (5.36,-2.57) -- (0,0) -- (5.36,2.57) -- cycle    ;
\draw    (250,125) -- (250,175) ;
\draw   (97,132) .. controls (97,125.37) and (102.37,120) .. (109,120) -- (328,120) .. controls (334.63,120) and (340,125.37) .. (340,132) -- (340,168) .. controls (340,174.63) and (334.63,180) .. (328,180) -- (109,180) .. controls (102.37,180) and (97,174.63) .. (97,168) -- cycle ;.48,180) and (97,175.52) .. (97,170) -- cycle ;
\draw   (208.33,132.5) .. controls (208.33,130.11) and (210.27,128.17) .. (212.67,128.17) -- (225.67,128.17) .. controls (228.06,128.17) and (230,130.11) .. (230,132.5) -- (230,145.67) .. controls (230,148.06) and (228.06,150) .. (225.67,150) -- (212.67,150) .. controls (210.27,150) and (208.33,148.06) .. (208.33,145.67) -- cycle ;
\draw   (109,160) -- (240,160) -- (240,176) -- (109,176) -- cycle ;

\draw (128.5,145) node    {$\mathbf{m} ,\boldsymbol{\tau }$};
\draw (172,132) node    {$F_{K}(\cdot)$};
\draw (211,135) node [anchor=north west][inner sep=0.75pt]    {$\mathbf{r}_{\boldsymbol{\tau }}$};
\draw (169.5,169) node    {$P:X\mapsto \mathbf{y}_{0} +X\cdot \mathbf{y}_{1}$};
\draw (253,140) node [anchor=west] [inner sep=0.75pt]    {$\mathbf{y}_{0} =\mathbf{m}$};
\draw (253,160) node [anchor=west] [inner sep=0.75pt]    {$\mathbf{y}_{1} =\left[\frac{\mathbf{r}_{\boldsymbol{\tau }} -\mathbf{m}}{\alpha }\right]_t$};

\end{tikzpicture}
    }
    \caption{{Polynomial Encoding. A message $\mathbf{m}\in \mathbb{Z}_t^N$ identified by $\boldsymbol{\tau}$ is encoded as $P$ using the secret $\alpha$ and the challenge vector $\textbf{\textit{r}}_{\boldsymbol{\tau}}$.}}
    \label{fig:vche2}
\end{figure}
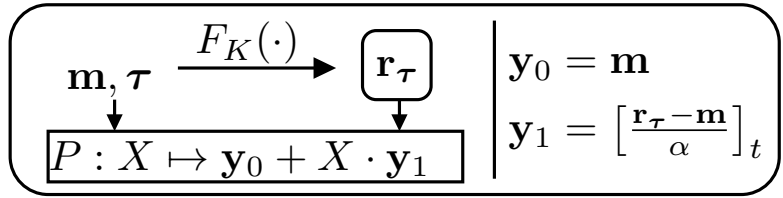

On input a vector $\mathbf{m}{\in} \mathbb{Z}_t^N$ with identifier $\boldsymbol{\tau}$, a secret key $\alpha$, and a pseudorandom function (PRF) $F_K(\cdot)$, the polynomial-based encoder outputs a polynomial $P$ such that $P(0){=}\mathbf{m}$ and $P(\alpha){=}\textbf{\textit{r}}_{\boldsymbol{\tau}}$, with $\textbf{\textit{r}}_{\boldsymbol{\tau}}$ the challenge vector obtained from the PRF $F_K(\cdot)$. Then, each component of $P$ is encoded using batching. A simplified version of the encoding is represented in Figure~\ref{fig:vche2} and the details of the encoder are shown in Scheme~\ref{ecd:PE}. 

\subsection{Polynomial-Based Authenticator}\label{sec:vche2:def}

We combine the polynomial-based encoding with an HE scheme and show that it instantiates a homomorphic authenticator~\cite{catalano2013practical}. Similar to REP, we describe the complete pipeline.

Let $F_K{:} \mathcal{I}{\rightarrow}\mathbb{Z}_t$ be a variable length PRF and $\text{HE}$ a secure homomorphic encryption scheme as in \S\ref{sec:prelim:fhe} with plaintext space $\mathcal{R}_t{=}\mathbb{Z}_t[X]/( X^N\!{+}1)$ of degree $N$. We define a \textit{Polynomial Encoding}-based authenticator (PE) in Scheme~\ref{scheme:PE} and briefly describe it here.

\descr{$\text{PE.KeyGen}(1^\lambda)$}$\rightarrow \!(\mathbf{evk}, \mathbf{sk})$: For $t$ a $\lambda$-bit prime number, it samples a random invertible point $\alpha$ in $\mathbb{Z}_t^*$. It sets up the PRF with key $K$ and the HE scheme such that both achieve at least $\lambda$-bit security. It generates the encryption and evaluation keys for the HE scheme. Finally, it outputs the authenticator evaluation and secret keys.

\descr{$\text{PE.Auth}(\mathbf{m}, \boldsymbol{\tau}; \mathbf{sk})$}$\rightarrow \boldsymbol{\sigma}$: On input a plaintext vector ${\mathbf{m}\in \mathbb{Z}_t^N}$ with identifiers ${\boldsymbol{\tau}}$ (and $N$ the degree of the plaintext polynomial ring), it calls \textbf{PolynomialEncoder}($\mathbf{m}, \boldsymbol{\tau}; \alpha, F_K(\cdot)$) that returns $(\mathbf{y}_0, \mathbf{y}_1){\in}\mathcal{R}_t^2$ (\underline{Encode}). Then, it encrypts $\mathbf{y}_0$ and $\mathbf{y}_1$ to $\mathbf{c}_0$ and $\mathbf{c}_1$ respectively using the HE scheme. It returns the authentication $\boldsymbol{\sigma}{=}(\mathbf{c}_0,\mathbf{c}_1)$. Note that initially, a message is encoded as a degree-one polynomial in $\mathcal{R}_q$ (\ie a length 2 list of polynomials) identified by its coefficients $\mathbf{c}_0$ and $\mathbf{c}_1$. Through the evaluation process, this degree can increase. We denote by $d$ the polynomial encoding's degree (\eg $d{=}1$ after a fresh authentication).

\begin{figure}[!t]
\refstepcounter{scheme}\label{ecd:PE}
\scheme{\textsc{\footnotesize{Polynomial-Based Encoder in}} \footnotesize{$\mathcal{R}_t{=}\mathbb{Z}_t[X]/( X^N\!\!{+}1)$}}{
~
$\textbf{PolynomialEncoder}(\mathbf{m}, \boldsymbol{\tau}; \alpha, F_K(\cdot))$:
\begin{enumerate}
    \item Compute $\mathbf{r}_{\boldsymbol{\tau}} {\in} \mathbb{Z}_t^N$ s.t. $\forall i {\in} [N]$, $\mathbf{r}_{\boldsymbol{\tau}}[i]{=}F_K(\boldsymbol{\tau}[i])$.
    \item Set $\mathbf{y}_0$ and $\mathbf{y}_1$ to be the batching encoding of $\mathbf{m}$ and  $[(\mathbf{r}_{\boldsymbol{\tau}}-\mathbf{m})\cdot \alpha^{-1}]_t$.
    \item Return $P=(\mathbf{y}_0, \mathbf{y}_1)$.
\end{enumerate}
}
\end{figure}

\descr{$\text{PE.Eval}(f(\cdot), \vv{\boldsymbol{\sigma}};\mathbf{evk})$}$\rightarrow \!\boldsymbol{\sigma}'$: A function $f(\cdot)$ represented by an arithmetic circuit $\mathcal{C}$ over the ciphertexts is evaluated gate by gate on the $n$ previously authenticated input vectors in $\vv{\boldsymbol{\sigma}}{=}(\boldsymbol{\sigma}_1, \!..., \boldsymbol{\sigma}_n)$. The output of each gate in $\mathcal{C}$ depends on its functionality: Additive gates execute the corresponding HE operation component-wise on the input authentications and multiplicative ones perform a convolution. Rotation gates execute the corresponding homomorphic rotation on each component of the input authentication. After the circuit evaluation, this procedure returns $\boldsymbol{\sigma}'$, the authentication output of the final gate in $\mathcal{C}$.

\descr{$\text{PE.Ver}(\mathcal{P}, \boldsymbol{\sigma}';\mathbf{sk})$}$\rightarrow\{0,1\}$: It parses the labeled program $\mathcal{P}{=}(f,(\boldsymbol{\tau}_1, $ $\!..., \boldsymbol{\tau}_n))$ and $\boldsymbol{\sigma}'{=} (\mathbf{c}_0, \!..., \mathbf{c}_d)$. Offline, it pre-computes the value $\boldsymbol{\rho}\!=\!\!f\!(\!F_K(\boldsymbol{\!\tau}_1\!),\!..., F_K(\boldsymbol{\!\tau}_n\!)\!)$. Online, it decrypts the encrypted vector $\boldsymbol{\sigma}'$ to $(\mathbf{y}_0, \!..., \mathbf{y}_d)$ and checks if $\boldsymbol{\rho} {=} \sum_{i=0}^{d} \mathbf{y}_i \cdot \alpha^i$. If the check passes, it accepts the result $\mathbf{y}_0$.

\begin{figure}[!t]
\refstepcounter{scheme}\label{scheme:PE}
\scheme{\textsc{\footnotesize{Polynomial Encoding-Based Authenticator}}}{
~
    \begin{itemize}[leftmargin=*]
    \itemsep0.7em
        \item $\textbf{PE.KeyGen}(1^\lambda)$: Let $t$ be a $\lambda$-bit prime number. 
        \begin{enumerate}[topsep=0pt, itemsep=0ex]
            \item Sample a random invertible point $\alpha {\leftarrow} \mathbb{Z}_t^*$. 
            \item Choose a PRF key $K {\leftarrow} \{0,1\}^*$ for at least $\lambda$-bits security.
            \item Initialise the HE keys $(\mathbf{pk}_{\text{HE}}, \mathbf{evk}_{\text{HE}}, \mathbf{sk}_{\text{HE}})\!\!=\!\! \text{HE.KeyGen}(1^{\lambda})$ for at least $\lambda$-bits security.
            \item Return $\mathbf{evk}\!\!=\!\!( \mathbf{evk}_{\text{HE}}, t)$ and $\mathbf{sk}\!\!=\!\!(K, \alpha, \mathbf{sk}_{\text{HE}})$.
        \end{enumerate}
        \item $\textbf{PE.Auth}(\mathbf{m}, \boldsymbol{\tau}; \mathbf{sk})$: For ${\mathbf{m}\in \mathbb{Z}_t^N}$ with identifiers in $\boldsymbol{\tau}$.
        \begin{itemize}[topsep=0pt, itemsep=0ex]
            \item \underline{Encode}: $(\mathbf{y}_0, \mathbf{y}_1) = \textbf{PolynomialEncoder}(\mathbf{m}, \boldsymbol{\tau}; \alpha, F_K(\cdot))$
            \item \underline{\smash{Encrypt}}:
        \begin{enumerate}[leftmargin=0pt, topsep=0pt, itemsep=0ex]
            \item Create $\mathbf{c}_i{=}\text{HE.Enc}(\mathbf{y}_i;\mathbf{pk}_{\text{HE}})$, $\forall i {\in} [0{:}1]$.
            \item Return $\boldsymbol{\sigma}{:=}(\mathbf{c}_0,\mathbf{c}_1)$.
        \end{enumerate}
        \end{itemize}
        \item $\textbf{PE.Eval}(f(\cdot), \vv{\boldsymbol{\sigma}}; \mathbf{evk})$: Let $\mathcal{C}$ be the HE arithmetic circuit of a function $f(\cdot)$ to be computed over previously authenticated encrypted inputs stored in $\vv{\boldsymbol{\sigma}}$.
        \begin{enumerate}[topsep=0pt, itemsep=0ex]
            \item Parse $\vv{\boldsymbol{\sigma}}{=}(\boldsymbol{\sigma}_1,\!..., \boldsymbol{\sigma}_n)$.
            \item Evaluate homomorphically each gate in $\mathcal{C}$ and output $\boldsymbol{\sigma}{=}(\mathbf{c}_0,\!...,\mathbf{c}_d)$, an authentication of degree $d$, s.t.:
                \begin{itemize}[leftmargin=-0.1cm, labelsep=0.05cm,]
                \item \underline{Addition}: On input two authentications $\boldsymbol{\sigma}_{1}$ and $\boldsymbol{\sigma}_{2}$ of degree $d_1$ and $d_2$ resp., set $d{=}\text{max}(d_1,d_2)$ and $\boldsymbol{\sigma}{=}\boldsymbol{\sigma}_{1}+\boldsymbol{\sigma}_{2}$, \ie $\forall k{\in}[d{+}1]$, $\boldsymbol{\sigma}[k]{=}\text{HE.Add}(\boldsymbol{\sigma}_{1}[k], \boldsymbol{\sigma}_{2}[k])$.
                \item \underline{\smash{Multiplication}}: On input $\boldsymbol{\sigma}_{1}$ and $\boldsymbol{\sigma}_{2}$ of degree $d_1$ and $d_2$ resp., set $d{=}d_1{+}d_2$ and perform a convolution, \textit{i.e.,} $\forall k{\in} [d{+}1]$, $\boldsymbol{\sigma}[k]{=}\sum_{i=1}^{k} \text{HE.Mul}(\boldsymbol{\sigma}_{1}[i], \boldsymbol{\sigma}_{2}[k{-}i{+}1]; \mathbf{evk}_{\text{HE}})$.
                \item \underline{\smash{Rotation by $r$}}: On input an authentication $\boldsymbol{\sigma}_{1}$ of degree $d_1$, set $d{=}d_1$ and rotate all components of $\boldsymbol{\sigma}$, \textit{i.e.,} $\forall k{\in}[d{+}1]$, $\boldsymbol{\sigma}[k]{=}\text{HE.Rot}_r(\boldsymbol{\sigma}_{1}[k]
                ; \mathbf{evk}_{\text{HE}})$.
                \end{itemize}
            \item Return $\boldsymbol{\sigma}'$, the authentication output of the final gate in $\mathcal{C}$.
        \end{enumerate}
    \item $\textbf{PE.Ver}(\mathcal{P}, \boldsymbol{\sigma}'; \mathbf{sk})$: Parse the target program $\mathcal{P} {=} (f,\allowbreak (\boldsymbol{\tau}_1, \!..., \boldsymbol{\tau}_n))$.
        Pre-compute the challenge values:
        \begin{enumerate}[topsep=0pt, itemsep=0ex]
            \item $\forall i {\in} [n]$, $\forall j {\in} [N]$, set $\mathbf{r}_{\boldsymbol{\tau}_i}{\in} \mathbb{Z}_t^N$ s.t. $\mathbf{r}_{\boldsymbol{\tau}_i}[j]{=}F_K(\boldsymbol{\tau}_i[j])$.
            \item Compute $\boldsymbol{\rho}{=}f(\mathbf{r}_{\boldsymbol{\tau}_1}, \!..., \mathbf{r}_{\boldsymbol{\tau}_n})$.
        \end{enumerate}
        Then, during the online phase, parse $\boldsymbol{\sigma}'{=}(\mathbf{c}_0, \dots, \mathbf{c}_d)$ obtained from the evaluation and:
        \begin{enumerate}[topsep=0pt, itemsep=0ex]
            \item Set the result to $\mathbf{y}_0 {=} \text{HE.Dec}(\mathbf{c}_0; \mathbf{sk}_{\text{HE}})$.
            \item Set $\mathbf{y}_i {=} \text{HE.Dec}(\mathbf{c}_i; \mathbf{sk}_{\text{HE}})$, ${\forall} i {\in}[0{:}d]$.
            \item Check if $\boldsymbol{\rho} \stackrel{?}{=} \sum_{i=0}^{d} \mathbf{y}_i \cdot \alpha^i$, else return $0$.
            \item Return $1$ (\ie accept).
        \end{enumerate}
    \end{itemize}
}
\end{figure}

\descr{Correctness.} The authentication correctness follows from the correctness of the HE scheme (\S\ref{sec:prelim:fhe}). The evaluation correctness follows from the construction. For linear gates (additions, multiplication by constant, and rotations), correctness follows by the correctness of the HE scheme encrypting the encoding of the input. The multiplicative gate is a convolution between the inputs: When executed on the ciphertext it ports to the plaintexts by the correctness of the HE scheme. By the definition of polynomial convolution, this leads to an encoding of the product of scalar inputs. PE differs from the original authenticator by Catalano and Fiore~\cite{catalano2013practical} by injecting, before encryption, the encoding in the native plaintext polynomial ring.

\descr{Security.} Within the scope of our threat model (\S\ref{sec:pb}), the following theorem states that a misbehaving server has only a negligible probability of cheating without being detected.

\begin{theorem}\label{th:pe}
If the PRF $F_K$ and the canonical HE scheme are at least $\lambda$-bit secure and if $t$ is a $\lambda$-bit prime number, then, for any program $\mathcal{P}$ with authentications of bounded degree, $\text{PE}$ is a secure authenticator and a PPT adversary has a probability of successfully cheating the verification negligible in $\lambda$.
\end{theorem}

\descrit{Proof Intuition.} The security follows from the security of the PRF and the polynomial identity lemma for various polynomials over the field $\mathbb{Z}_t$ (as $t$ is a prime number); the probability of the adversary cheating without being detected is negligible in $\lambda$. The formal proof is presented in Appendix~\ref{ap:vche2}.

\descr{Overhead.} After the evaluation, the size of the resulting PE authentication $\boldsymbol{\sigma}'$ grows polynomially with the degree of the polynomial computation due to the convolution between the authentications. Hence, for programs with large multiplicative depth, PE might introduce a significant communication overhead between the client and the server and become less communication efficient than REP. This growth can also affect the client's computational overhead due to the cost of pre-computing the challenges plus the cost of the polynomial evaluation (Step {\small3} in PE.Ver). To account for this issue, we design a new \textit{polynomial compression protocol} (\S\ref{sec:vche2:PP}). Moreover, the convolutions introduced by the multiplicative gates during the evaluation phase yield a non-negligible computational and communication overhead for the computing server (see \S\ref{sec:system-impl}). For client-aided scenarios, we mitigate the effects on the computation and communication overheads, by introducing a new technique called \textit{re-quadratization} (\S\ref{sec:vche2:RQ}).

\subsection{Polynomial Compression Protocol}\label{sec:vche2:PP}
\begin{figure}[!t]
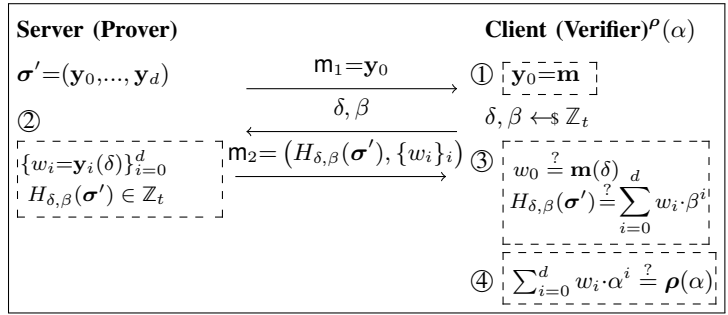

    \procbox[width=0.35\textwidth, colspace=-0.2cm]{}{\\[-0.7\baselineskip]
        \textbf{Server (Prover)} \>  \> \textbf{Client (Verifier)}^{\text{$\boldsymbol{\rho}$}}\text{$(\alpha)$}\\[-0.2\baselineskip]
        \text{$\boldsymbol{\sigma}' {=} (\mathbf{y}_0, \!..., \mathbf{y}_d)$} \> \sendmessageright*[2.8cm]{\text{$\textsf{m}_1 {=} \mathbf{y}_0$}} \> \hspace{-0.2cm}\raisebox{.5pt}{\textcircled{\raisebox{-.8pt} {\small 1}}}~\dbox{\text{$\mathbf{y}_0 {=} \mathbf{m}$} }\\[-0.7\baselineskip]
        \text{\raisebox{-1pt}{\textcircled{\raisebox{-.8pt} {\small 2}}}} \> \sendmessageleft*[2.8cm]{\text{$\delta, \beta$}} \> \text{$\delta$}, \text{$\beta$} \sample \text{$\mathbb{Z}_t$}\\[-0.7\baselineskip]
      \dbox{\hspace{-0.2cm}
        \begin{subprocedure}
            \procedure{}{
           \text{$\!\!\{w_i {=} \mathbf{y}_i(\delta)\}_{i=0}^d$}\\
            \text{$H_{\delta,\beta}(\boldsymbol{\sigma}') \in \mathbb{Z}_t$}
            }
        \end{subprocedure}
        } \> \sendmessageright*[2.8cm]{\text{\,\,\,$\textsf{m}_2{=}\left(H_{\delta, \beta}(\boldsymbol{\sigma}'),  \{w_i\}_i\right)$}} \>  
       \hspace{-0.2cm}\raisebox{.5pt}{\textcircled{\raisebox{-.8pt} {\small 3}}}~\dashbox[8.7em][l]{
       \begin{subprocedure}
            \procedure{}{
                \hspace{-0.2cm} \text{$w_0 \stackrel{?}{=} \mathbf{m}(\delta)$} \\[-0.3cm]
                \hspace{-0.2cm} \text{$\!H_{\delta,\beta}(\boldsymbol{\sigma}')\!\stackrel{?}{=} \!\sum_{i=0}^d w_i {\cdot} \beta^i$\hspace{0.2cm}}
            }
        \end{subprocedure}
        }\\[-0.\baselineskip]
        \> \> \hspace{-0.2cm}\raisebox{.5pt}{\textcircled{\raisebox{-.8pt} {\small 4}}}~\dbox{
        \text{\hspace{-0.1cm}$\sum_{i=0}^d w_i {\cdot} \alpha^i\stackrel{?}{=} \boldsymbol{\rho}(\alpha)$\hspace{-0.2cm}}
        }
    }
    \caption{{Polynomial Compression Protocol (PoC, \S\ref{sec:vche2:PP}). For clarity, polynomials are represented in the plaintext space.}}\label{proto:poly}
\end{figure}

As seen in \S\ref{sec:vche2:def}, the size of the PE authentication $\boldsymbol{\sigma}'$ grows linearly with the degree of the function represented by the evaluated circuit (\ie it comprises more and more ciphertexts). This can cause a significant communication overhead between the server and the client. To mitigate this overhead, we design a polynomial compression protocol (PoC) that compresses the authentication from $d{+}1$ ciphertexts to only two, for authentications of degree $d$. 
Informally, once the computing server (\ie the prover) has sent the claimed output $\mathbf{y}_0$, the client (\ie the verifier) challenges the server with a hash function. The server responds with the hash digest and the intermediate results $\{w_i\}_{i=0}^d$. A sketch of our PoC can be seen in Figure~\ref{proto:poly}. This protocol is made non-interactive in the ROM using the Fiat-Shamir heuristic.  
As we will see in \S\ref{sec:eval}, the PoC reduces the communication and verifier overhead at the cost of higher runtimes for the prover. 

The combination of PE with PoC leads to a secure authenticator. Informally, the compression uses a polynomial hashing mechanism that by the polynomial identity lemma gives a negligible soundness error for an appropriate choice of parameters. We provide the full proof in Appendix~\ref{ap:PoC}.

\subsection{Interactive Re-Quadratization}\label{sec:vche2:RQ}
As discussed in \S\ref{sec:vche2:def}, PE executes a convolution for every multiplication gate of the evaluation circuit. Hence, due to the growing number of homomorphic multiplications and ciphertexts, the longer the input encodings are, the heavier the server's operations are. 
When interactions between the server and the client are possible (\ie in client-aided settings), we propose a \textit{re-quadratization} (ReQ) (similar to the relinearization used in HE). 
It converts an authentication comprising five ciphertexts (\ie $\boldsymbol{\sigma}' {=} (\mathbf{c}_0, \!..., \mathbf{c}_4)$ for the authentication of a depth-two computation) to one of three ciphertexts (\ie related to a depth-one). 
In a nutshell, ReQ evaluates part of the verification procedure for the terms of higher degree (\ie $\mathbf{c}_3$ and $\mathbf{c}_4$) and embeds the result in the lower terms (\ie $\mathbf{c}_1$ and $\mathbf{c}_2$). 
Our interactive protocol bounds the authentication to, at most, three ciphertexts, thus limiting the computational burden of computing the convolution of high-degree polynomials for the server (see \S\ref{sec:eval}).

\begin{figure}[!t]
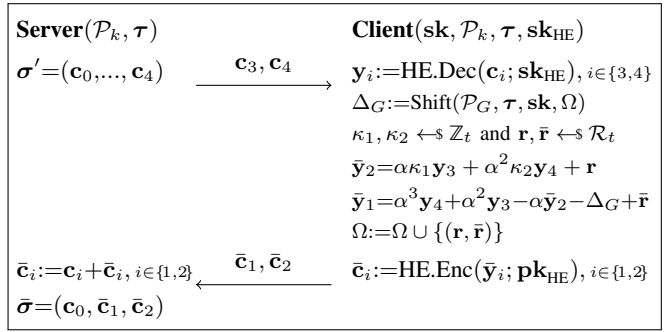

    \procbox[width=0.33\textwidth, colspace=-0.3cm]{ }{\\[-0.7\baselineskip]
        \textbf{Server}(\mathcal{P}_k, \boldsymbol{\tau}) \>  \> \textbf{Client}(\mathbf{sk}, \mathcal{P}_k, \boldsymbol{\tau}, \mathbf{sk}_{\text{HE}})\\[-0.3\baselineskip]
        \boldsymbol{\sigma}' {=} (\mathbf{c}_0, \!..., \mathbf{c}_4) \> \sendmessageright*[1.8cm]{\mathbf{c}_3, \mathbf{c}_4} \> \mathbf{y}_i {:=} \text{HE.Dec}(\mathbf{c}_i;\mathbf{sk}_{\text{HE}}), {\scriptstyle i{\in}\{3,4\}} \\[-0.5\baselineskip]
       \> \> 
        \begin{subprocedure}
            \procedure{}{
            \Delta_G {:=} \text{Shift}(\mathcal{P}_G, \boldsymbol{\tau}, \mathbf{sk}, \Omega)\\
            \kappa_1, \kappa_2 \sample \mathbb{Z}_t \text{ and } \mathbf{r}, \bar{\mathbf{r}} \sample \mathcal{R}_t \\
            \bar{\mathbf{y}}_2{=}\alpha\kappa_1 \mathbf{y}_3 + \alpha^2\kappa_2 \mathbf{y}_4 + \mathbf{r} \\
            \bar{\mathbf{y}}_1{=} \alpha^3 \mathbf{y}_4 {+} \alpha^2 \mathbf{y}_3 {-} \alpha \bar{\mathbf{y}}_2 {-} \Delta_G {+} \bar{\mathbf{r}}\\
            \Omega {:=} \Omega \cup \{(\mathbf{r}, \bar{\mathbf{r}})\} 
            }
        \end{subprocedure} \\[-0.3\baselineskip]
       \bar{\mathbf{c}}_i {:=} \mathbf{c}_i {+} \bar{\mathbf{c}}_i,  {\scriptstyle i{\in}\{\!1,2\!\}}\,\,\> \sendmessageleft*[1.8cm]{\bar{\mathbf{c}}_1, \bar{\mathbf{c}}_2} \> \bar{\mathbf{c}}_i {:=} \text{HE.Enc}(\bar{\mathbf{y}}_i;\mathbf{pk}_{\text{HE}}), {\scriptstyle i{\in}\{\!1,2\!\}}\\[-0.4\baselineskip]
       \bar{\boldsymbol{\sigma}} {=} (\mathbf{c}_0, \bar{\mathbf{c}}_1, \bar{\mathbf{c}}_2) \>\>
    }
    \caption{{Interactive Re-Quadratization (ReQ) for gate $G$ (\S\ref{sec:vche2:RQ}).}}\label{proto:RQ}
\end{figure}

Figure~\ref{proto:RQ} presents our protocol in detail. In a nutshell, the client decrypts the higher term ciphertexts sent by the server (\ie $\mathbf{c}_3$ and $\mathbf{c}_4$) and returns to the server encryptions of two masked random linear combinations of the plaintexts. The random masks create an offset of the encoding that is removed by the shift value $\Delta_G$ that keeps track of the offset up to the re-quadratization after the $G$-th multiplication gate. The correctness of PE combined with ReQ follows directly by construction. The security of the combination is guaranteed by the random maskings. A full description of ReQ and its security proof are presented in Appendix~\ref{ap:ReQ}.

Similar to other client-aided operations~\cite{juvekar2018gazelle,akavia2022achievable,akavia2018secure}, ReQ trades-off interactivity for higher performance. The theoretical communication complexity of ReQ is the exchange of four ciphertexts per re-quadratization. This leads to a linear overhead in the degree of the function as in the original PE. The memory complexity for the client is to keep the state of the blinding list $\Omega$ (storing two new polynomials per re-quadratization). The computation complexity for the client is to execute two HE encryption/decryption operations and to compute the offset $\Delta_G$ (\ie a plaintext convolution). Overall, ReQ requires that (i)~both client and server be online, and (ii)~the client engages in some computations. Nevertheless, ReQ has the major advantage of maintaining a $\boldsymbol{\sigma}'$ of at most degree-two, which implies (i)~lower communication overhead to obtain the result, and (ii) lower computation overhead for the server compared to the original PE (as the convolution now involves at most two degree-two polynomials).

\section{VERITAS}\label{sec:system-impl}
We introduce \namecomma, a new library that implements the two encodings and authenticators (and their optimizations) described in \S\ref{sec:vche1} and \S\ref{sec:vche2}. We first present our implementation (\S\ref{sec:impl}), before benchmarking native HE operations (\S\ref{sec:benchmark}) and evaluating it over several use-cases (\S\ref{sec:eval}). We then compare \namecomma' performance with prior work (\S\ref{sec:eval:prior}) and summarize the main takeaways (\S\ref{sec:eval:summ}).

\subsection{Implementation and Hardware}\label{sec:impl}

\name facilitates the transition from a standard HE pipeline (Figure~\ref{fig:fhepure}) to a pipeline enhanced with computation verification capabilities (Figure~\ref{fig:fhe}) by implementing (i) our two encoders and (ii) the corresponding authenticators. \name instantiates REP and PE with the BFV homomorphic encryption scheme, and Blake2b~\cite{rfc7693} as a PRF. \name offers flexible parameterization of the authenticator security parameter ($\lambda$) to provide sufficient security guarantees, depending on the application requirements (typically, $\lambda {\geq} 32$ for malicious-but-rational adversary models~\cite{lindell2013fast}), and it sets the HE and hash function security requirements to $\text{max}(\lambda,128)$-bits. By default, using \name does not provide any feedback to the computing server (see Appendix~\ref{ap:disc:outcome}). As BFV is a semantically secure canonical homomorphic encryption scheme and Blake2b yields a secure PRF~\cite{luykx2016security}, and as both are parameterized to achieve $128$-bits security, \name securely instantiates REP and PE. Thus, it protects the privacy of the client data and the computation result; by semantic security of BFV, the ciphertexts and the authentications reveal nothing about the underlying data before decryption. We build \name in Golang on top of Lattigo's BFV implementation~\cite{lattigo220}. It is modular and enables developers to seamlessly obtain client verification capabilities in existing homomorphic encryption pipelines (\ie by only changing a few lines of code). \name can be employed to verify the result correctness of any circuit admissible by BFV. We show its versatility over various use cases in \S\ref{sec:eval}. All experiments were conducted on a machine with an Intel Xeon E5-2680 v3 processor and 256GB of RAM. 

\subsection{Benchmarking BFV Operations}\label{sec:benchmark}
As described in \S\ref{sec:vche1} and \S\ref{sec:vche2}, both authenticators trivially support the BFV linear operations (\eg addition, subtraction, constant multiplication): These operations are simply executed on all components of the authentication $\boldsymbol{\sigma}$. Although both authenticators support any rotations, REP requires keys with an increase of the rotation step by a factor of $\lambda$ to avoid mixing up the various slots. These linear operations do not expand the size of the authentication. REP naturally supports multiplication operations, whereas PE requires a convolution for every multiplication. The computation complexity of this operation depends on the degree $d$ of the input authentication. Both REP and PE naturally support relinearization. 
We present benchmarks of the amortized evaluation costs (over the ciphertext slots) for each BFV operation in Table~\ref{table:bench} for $\lambda{=}32$, averaged over 100K runs. The BFV parameters are set to ($\log N{=}14$, $\log q{=}438$). We observe that REP's linear operation timings are multiplied by $\lambda$ (as the messages are replicated $\lambda$ times), whereas PE less than triples them. Regarding multiplications, REP's computational overhead remains constant and the computational overhead of the PE scheme increases with every depth of the circuit. Furthermore, each multiplication introduces a linear growth of the authentication size hence affects memory and communication. In terms of memory requirements, REP has in the worst case a linear overhead in $\lambda$ (this can be reduced if the whole encoding fits into a single ciphertext). For PE, the size $d$ of the authentication is linear in the degree of the evaluated function. We discuss in Appendix~\ref{ap:overhead} the influence of the security parameter $\lambda$ on \namecomma' overhead.

\begin{table}
\footnotesize
\centering
\caption{{Amortized timings of \name for homomorphic operations evaluation ($\mu$s) for an authenticator's security parameter $\lambda{=}32$. The baseline is the standard BFV scheme.}}
\begin{tabular}{ >{\centering\arraybackslash\columncolor[gray]{0.9}}L{0.6cm}|C{.6cm}>{\centering\arraybackslash\columncolor[gray]{0.9}}C{.8cm}C{.6cm}>{\centering\arraybackslash\columncolor[gray]{0.9}}C{.7cm}C{.6cm}>{\centering\arraybackslash\columncolor[gray]{0.9}}C{.6cm}C{.6cm}}
 Op.  & Add. & Mul. & Rot. & Relin. & \multicolumn{3}{c}{Mul. depth}\\
  & & const. &  &  & 1 & 2 & 3 \\
  \hline
 BFV  & $0.02$ & $0.04$ & $2.3$ & $1.8$ & $2.5$ & $3.5$ & $4.0$ \\
 \hline
 REP   & $1.07$ & $1.4$ & $70$ & $60$ & $105$ & $108$ & $118$ \\
 \hline
 PE    & $0.09$ & $0.1$ & $2.9$ & $4.8$ & $9.3$ & $15.0$ & $30.1$ \\
\end{tabular}
\label{table:bench}
\end{table}

\subsection{Experimental Case Studies}\label{sec:eval}
\begin{figure*}[!t]
     \centering
     \begin{subfigure}[b]{0.49\textwidth}
         \centering
         \includegraphics[width=\textwidth]{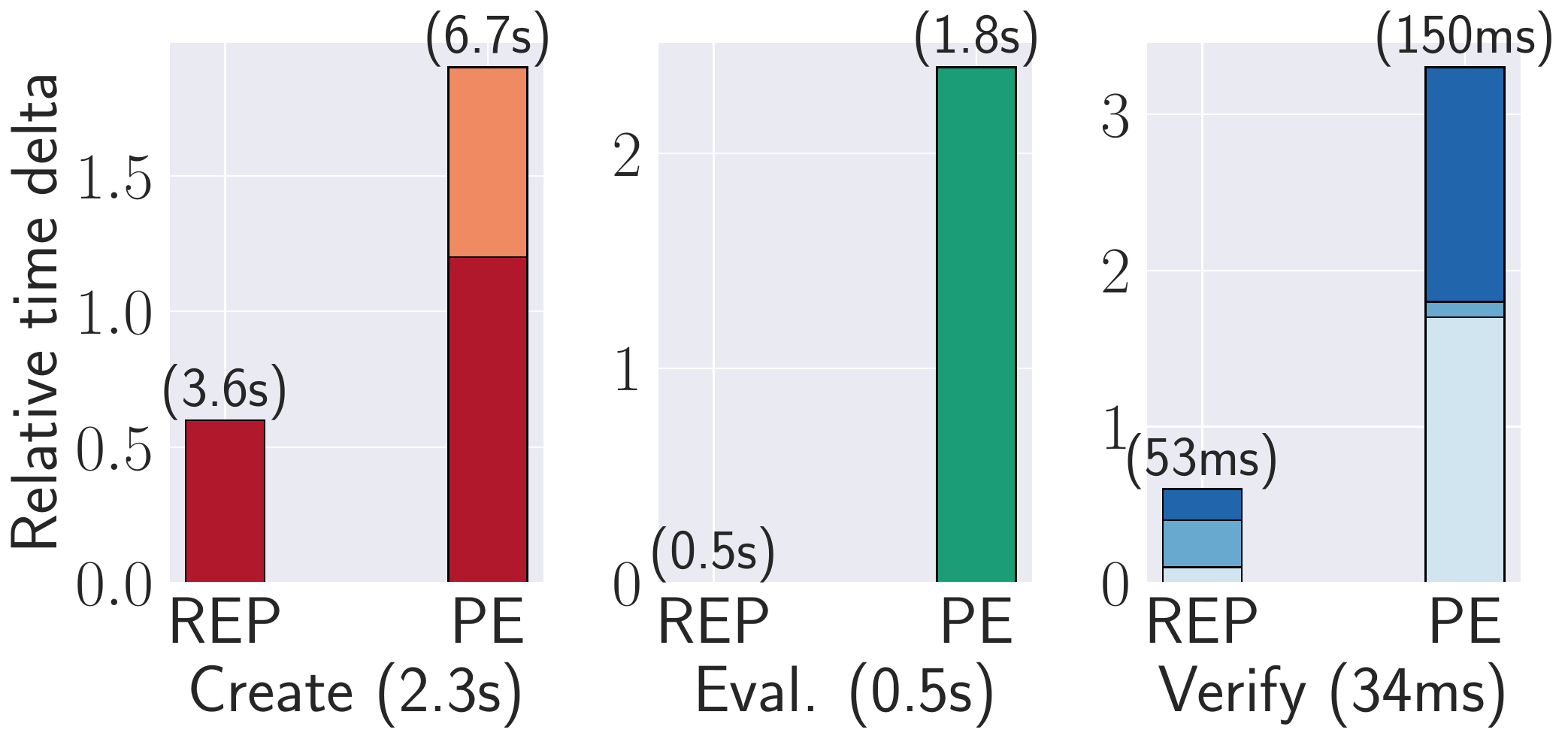}
         \caption{Ride-Hailing}
         \label{fig:OR}
     \end{subfigure}
     \hfill
     \begin{subfigure}[b]{0.49\textwidth}
         \centering
         \includegraphics[width=\textwidth]{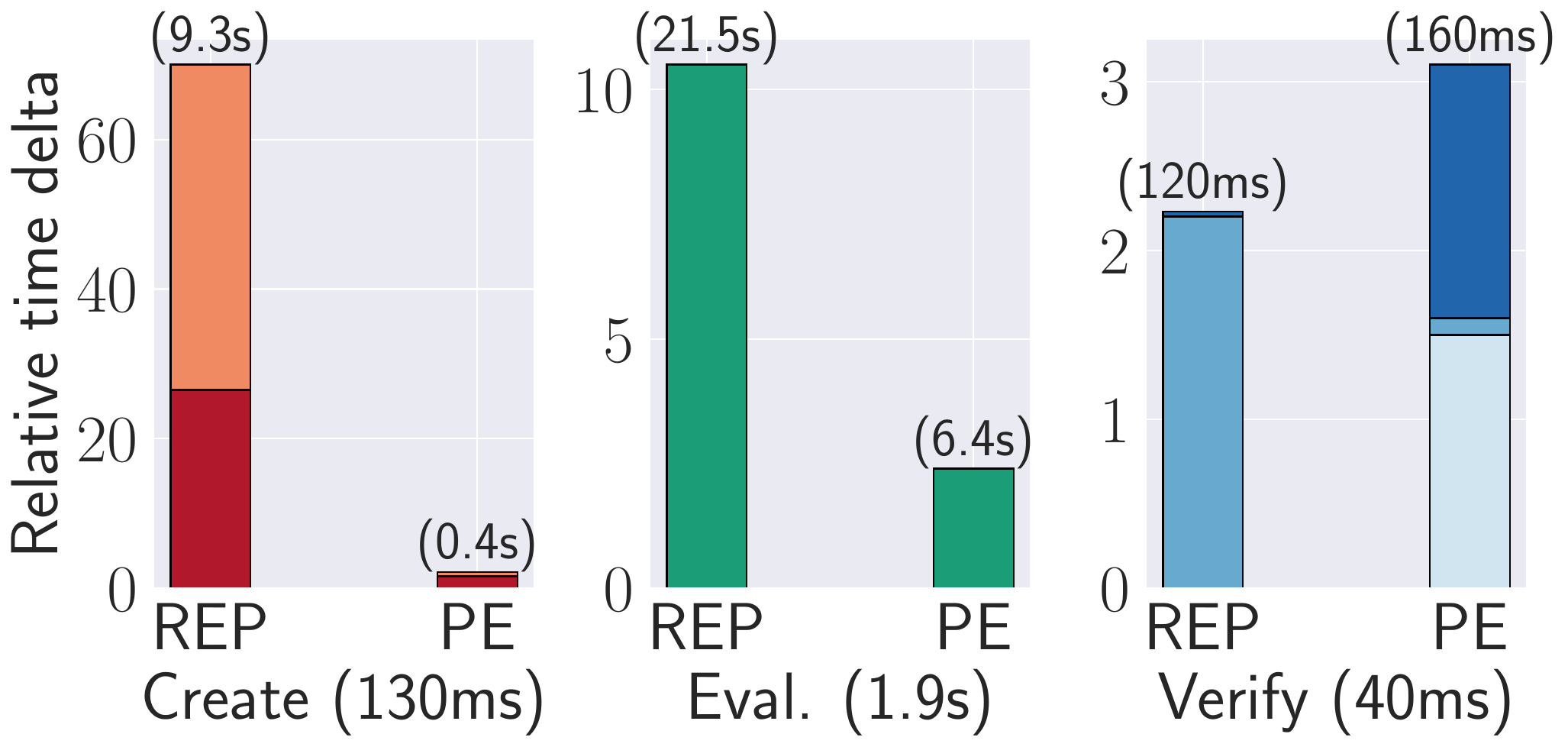}
         \caption{Disease Susceptibility}
         \label{fig:DS}
     \end{subfigure}
     \hfill
     \begin{subfigure}[b]{0.49\textwidth}
         \centering
         \includegraphics[width=\textwidth]{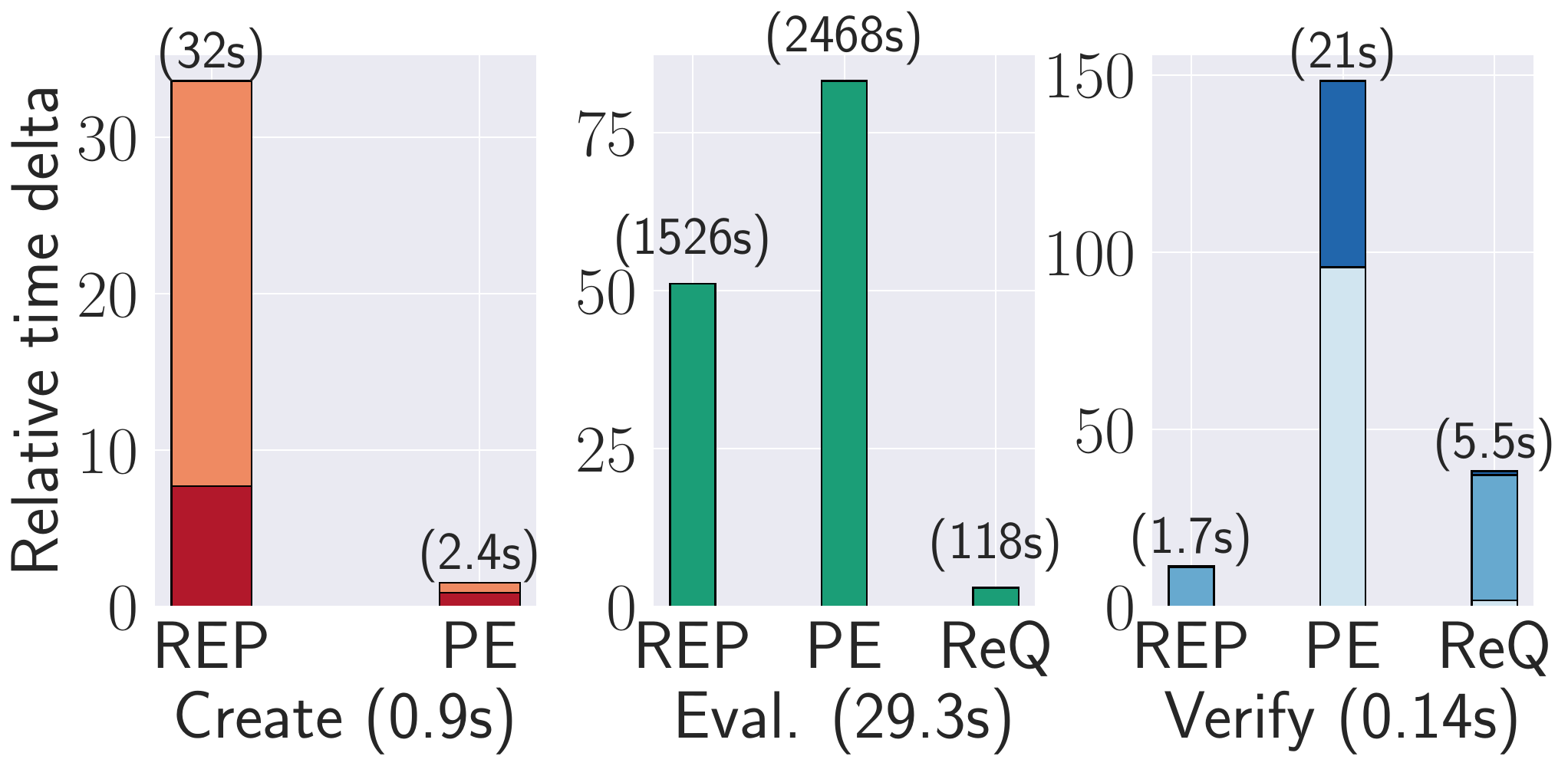}
         \caption{DNS Lookup}
         \label{fig:DNS}
     \end{subfigure}
     \hfill
     \begin{subfigure}[b]{0.49\textwidth}
         \centering
         \includegraphics[width=\textwidth]{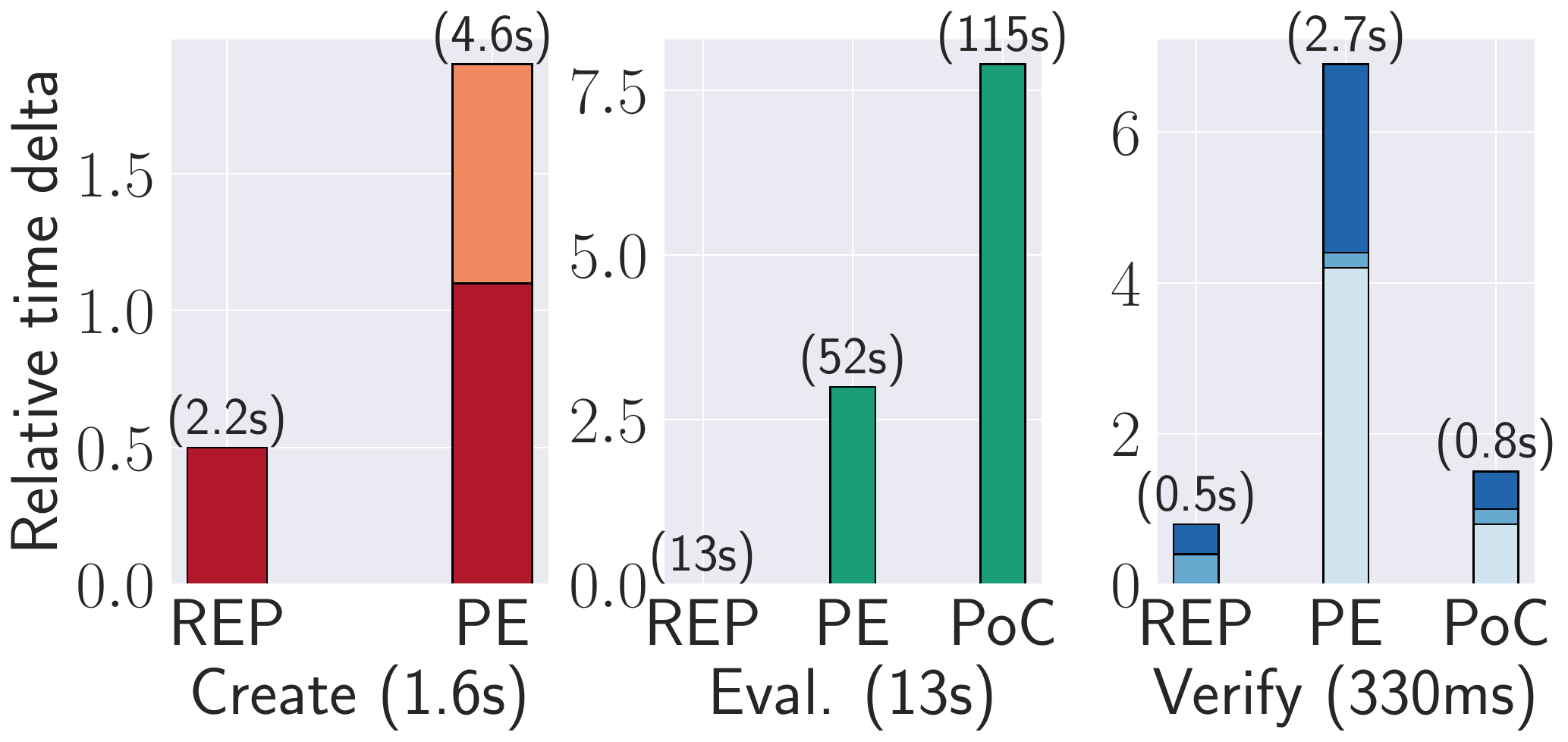}
         \caption{ML inference}
         \label{fig:lola}
     \end{subfigure}
     \hfill
     \begin{subfigure}[b]{0.49\textwidth}
         \centering
         \includegraphics[width=\textwidth]{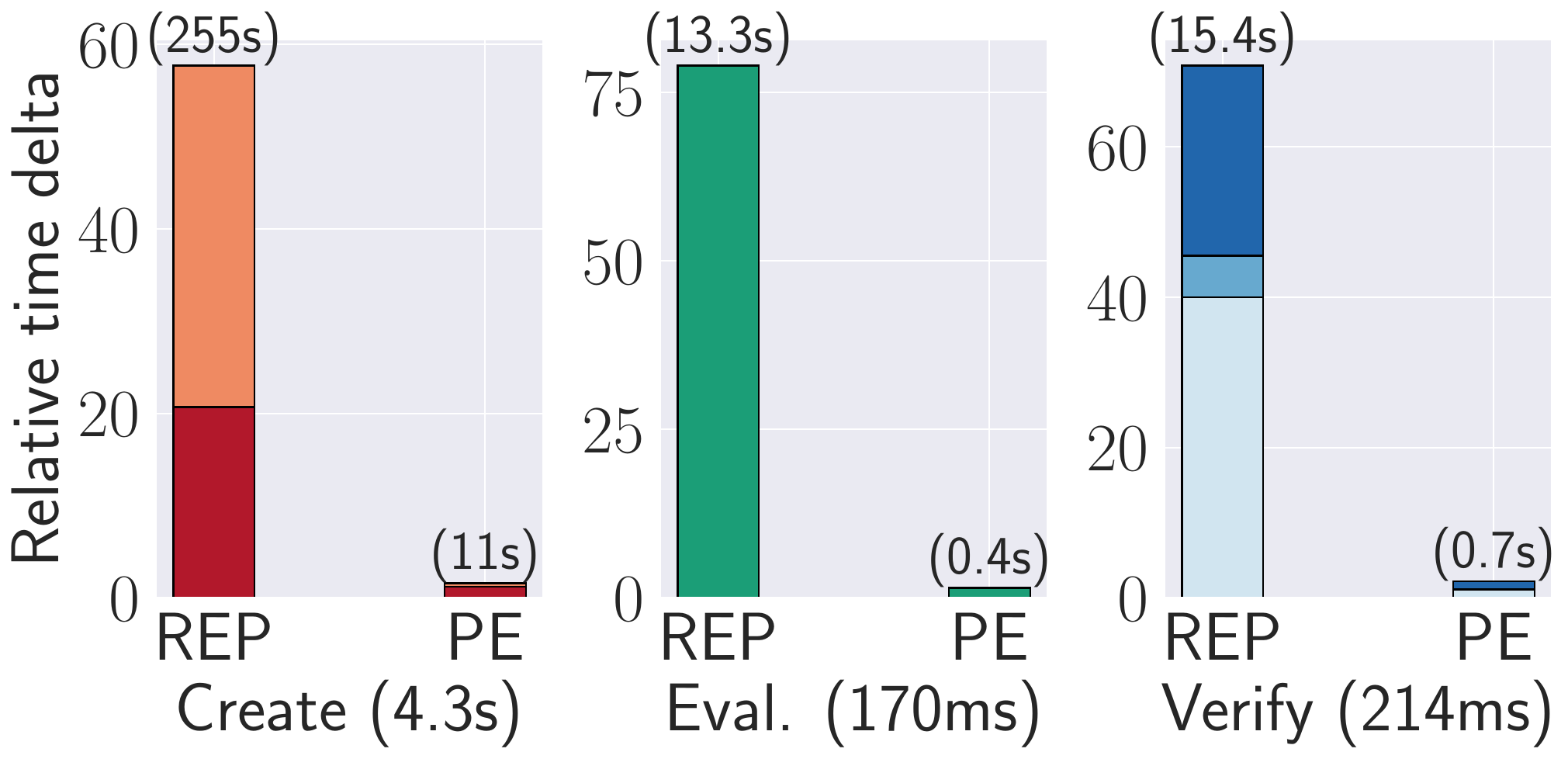}
         \caption{FedAvg per epoch}
         \label{fig:FL}
     \end{subfigure}
     \begin{subfigure}[b][][b]{0.1\textwidth}
         \centering
         \includegraphics[width=\textwidth]{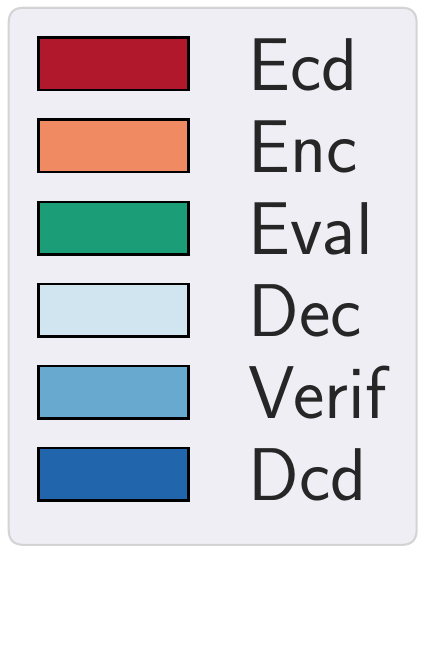}
     \end{subfigure}
     \hfill
    \begin{subtable}[b]{0.38\textwidth}
         \centering
\begin{tabular}{ c|>{\centering\arraybackslash\columncolor[gray]{0.9}}cc>{\centering\arraybackslash\columncolor[gray]{0.9}}cc}
& \rot{\small{REP}} & \rot{\small{PE}} & \rot{\small{PoC}} & \rot{\small{ReQ}} \\
\hline
(a) & 0/0 & 1/2 & - & - \\
(b) & 63/63 & 1/2 & - & -  \\
(c) & 63/0 & 1/130 & - & 14+1/14+2 \\
(d) & 0/0 & 1/3 & -/ 1 & - \\
(e) & 55/55 & 1/1 & - & - \\
\end{tabular}
         \caption{Relative communication delta Sent/Received per client, for each use-case.}
         \label{tab:com}
     \end{subtable}
        \caption{{\namecomma' computation overhead (1K runs avg.) for various use-cases (Figs.~\ref{fig:OR}-\ref{fig:FL}). The x-axis displays the authenticator/optimization employed. The y-axis represents the relative time delta wrt the HE baseline. The values in parenthesis on top of the bars represent the actual execution times. The values in parenthesis after each stage ({\scriptsize Create, Eval., Verify}) correspond to the HE baseline runtimes. PoC and ReQ are PE's optimizations from \S\ref{sec:vche2:PP} and \S\ref{sec:vche2:RQ}, resp. The various operations are the client encoding {\scriptsize(Ecd)} and encryption {\scriptsize(Enc)}, the server evaluation {\scriptsize(Eval)}, the client decryption {\scriptsize(Dec)}, verification of the challenges {\scriptsize(Verif)}, and decoding {\scriptsize(Dcd)}. ReQ and PoC overheads are included in the {\scriptsize(Eval)} and {\scriptsize(Verif)} stages. For each use-case, Table~\ref{tab:com} displays \namecomma' relative communication overhead w.r.t. the HE baseline.}}
        \label{fig:bench}
\end{figure*}

We now show how \name introduces computation verification capabilities in existing homomorphic encryption pipelines. We illustrate the performance of \name under different conditions: small (ride-hailing \S\ref{sec:eval:rh}), genomic-data analysis \S\ref{sec:eval:genomics}, federated averaging \S\ref{sec:eval:fl}) vs. large (encrypted search \S\ref{sec:eval:search}) multiplicative depth, and small (\S\ref{sec:eval:genomics}, machine-learning prediction \S\ref{sec:eval:inference}) vs. large (\S\ref{sec:eval:search} and~\ref{sec:eval:fl}) amount of data.
For each use case, we first describe the baseline (\ie the homomorphic encryption pipeline that protects only privacy) and then analyze the performance of \name using both authenticators and their optimizations when relevant. Note that, although the PE operations are embarrassingly parallelizable, we evaluate them on a single thread for the sake of comparison. \namecomma' relative (with respect to the baseline) computation and communication overheads (\ie $(x-x_{\text{base}})/x_{\text{base}})$) are presented in Figure~\ref{fig:bench}; these overheads are averaged over 1K runs. We group the homomorphic authenticator (HA) procedures into three stages: (a) the \textbf{Create} stage represents the HA.KeyGen() and HA.Auth() procedures executed by the (offloading) client, (b) the \textbf{Eval.} stage that accounts for the HA.Eval() ran by the computing server, and (c) the \textbf{Verify} stage which invokes the (decrypting) client HA.Ver() procedure.

\subsubsection{Ride-Hailing Services}\label{sec:eval:rh} 

Ride-hailing services enable the matching of a driver to a customer (or rider), depending on their locations. As location data can leak sensitive information~\cite{hern_fitness_2018,nytimes_lastnight_2018}, existing solutions use HE to protect confidentiality against the ride-hailing service~\cite{aivodji2018sride,pham2017oride}. Furthermore, it is crucial to verify the correctness of the matching in order to ensure a fair and transparent process~\cite{bokanyi2020understanding,li2020privacy}.

\descr{Description.} We evaluate an HE pipeline inspired by ORide~\cite{pham2017oride}. Each driver encrypts their location into a single ciphertext (each coordinate into a different slot piloted by the driver ID). The rider encrypts its location into a fully packed ciphertext (replicating its coordinates through the whole vector). The server computes the squared difference between the ciphertext of the rider and the ciphertext resulting from the sum of all drivers' ciphertexts. It returns the result to all drivers and to the rider who decrypts to discover the closest match. Using \namecomma, the drivers and the rider collectively run the authenticator-key generation and encode, before encryption, their respective location vectors.

\descr{Parameterization.} 
In our use-case, we consider 32 drivers, and set the HE cryptographic parameters to $\log N{=}15$ and $\log q{=}700$. 
With BFV, the client and drivers need in total $2.3$s to encrypt their location, and the server's matching requires $0.5$s. The client decryption/decoding needs $34$ms and the client egress/ingress communication per client is $6.3$MB. 
We parameterize the encodings with a security parameter $\lambda \geqslant 40$ (\ie $\lambda{=}64$ for REP and $t$ a $56$-bit prime for PE). Due to this use-case's low multiplicative depth, we do not present the PoC and ReQ optimizations.

\descr{Results.} 
In Figure~\ref{fig:OR}, we observe that both authenticators induce minimal overheads for the offloading client. With REP, the client's computation time for the creation of the encodings is multiplied by $1.6$, whereas PE is $1.9\times$ slower than the BFV baseline for encoding and encryption. This modest overhead is due to the small amount of data handled: REP's extended vector fits into a single ciphertext. As a result, the client needs $3.6$s and $6.7$s to generate, respectively, a REP and a PE authentication. REP has almost no effect on the evaluation time, whereas the circuit's multiplication leads to a $2.4\times$ increase in the server evaluation with PE ($1.8$s). The effort of the decrypting client for REP is mainly for the verification of the challenges, whereas for PE this is negligible; its cost is dominated by the decryption and decoding of the ciphertexts. Indeed, compared to the baseline ($34$ms), it takes $3.3\times$ more time for PE to decrypt, decode, and verify the result of the challenges. Although REP does not affect the communication overhead between the server and the client, PE doubles the server's ingress and triples the egress costs (see Table~\ref{tab:com}).

\subsubsection{Genomic-Data Analysis}\label{sec:eval:genomics} 

The emergence of direct-to-consumer and medical services that collect DNA to improve users' health and to customize their treatment~\cite{23andme_dna_nodate,dnafit}, along with the immutable and personal nature of genomic data, raises numerous privacy concerns~\cite{erlich2014routes}. As a result, several works consider protecting the confidentiality of the genomic data with HE~\cite{ayday_protecting_2013,danezis2014fast,decristofaro2013secure,wang_healer:_2016}. The medical nature of the computations performed on this data entails the correctness of the outcome to avoid misdiagnosis or insurance fraud~\cite{backes2015adsnark,turkmen2016igenopri}.

\descr{Description.} We apply \name to a disease-susceptibility computation. Users offload an encrypted version of their DNA (single nucleotide polymorphisms or SNPs) to the server. Later on, a medical institution offloads encrypted weights to the server; these are used for a specific disease prediction. The server computes the scalar product between the weights and the SNPs, and it returns the result to the user. The user can verify the result, even if it does not have access to the weights in cleartext and has access only to their identifiers shared by the medical institution. 

\descr{Parameterization.} The user encrypts $2^{15}$ SNPs of her DNA, and the medical center encrypts their corresponding weights for breast cancer prediction~\cite{gwas}. The HE cryptographic parameters are thus set to ($\log N{=}15$, $\log q {=}700$). The baseline client encoding/encryption time is $0.13$s, and the disease susceptibility computation by the server requires $1.9$s. The decryption/decoding needs $39$ms, and the server ingress/egress communication is $6.3$/$6.3$MB, respectively. We set the authenticator security parameters to $\lambda{=}64$ for REP and to $\log t{=}56$ for PE.

\descr{Results.} Figure~\ref{fig:DS} shows that, as the client offloads a fully packed ciphertext, the overhead induced by REP's creation is $~70\times$ the baseline ($9.2$s). In contrast, PE only triples the creation time ($0.4$s). The ingress communication to the server and its computations abide to a similar scaling ($404$MB for REP and $13$MB for PE). The server evaluation time, compared to the baseline, is tripled in the case of PE ($6$s) and is $10\times$ for REP ($22$s). This is due to an optimized evaluation circuit for REP that only performs the required inner sum after the 64 SNPs/weights ciphertexts are multiplied slot-wise; a naive evaluation would lead to a $\lambda$-factor overhead. The ingress communication to the decrypting client is tripled for PE, whereas it remains unchanged for REP. We also observe that REP's verification time ($123$ms) is dictated by the verification of the challenges ($\lambda/2$ of them), whereas the fast verification time of PE's challenge is shadowed by the decryption and decoding of $2$ additional ciphertexts compared to the baseline (leading to $157$ms).

\subsubsection{Search on Encrypted Data}\label{sec:eval:search} 

Encrypted lookups protect the confidentiality of sensitive data while enabling clients to query a database that stores it~\cite{akavia2018secure,choi2021compressed,roy2017hardware,wen2020leaf}. As both the database and the query are encrypted, the executing server does not learn any information. This can be applied, for instance, to an encrypted DNS search or to a query for the existence of a password in a list of vulnerable passwords~\cite{naor2019not}. In such cases, an unencrypted query would leak information to the server, \eg the websites a client is trying to access or their passwords; and an incorrect search result could lead to an application-level vulnerability, \eg redirection to a malicious website, or to the use of an insecure password.

\descr{Description.} We implement an encrypted DNS search in a database of domains of at most 16 characters in ASCII bit-representation. We follow the blueprints of the lookup use-case implemented in HElib~\cite{HELib} that we adapt to BFV. The query's bit-representation is XORed with the bit-representation of all database entries. If any of the result's bits are \textit{null} (\ie this entry did not match the query), its bit representation is multiplied by 0. The results for each entry are aggregated into a single ciphertext. For efficiency, several database entries are packed into a single ciphertext. We deliberately choose this use-case to demonstrate the impact of a large multiplicative depth on \namecomma' efficiency.

\descr{Parameterization.} The authenticator parameters are set to $\lambda{=}64$ for REP and to $\log t{=}58$ for PE. To support the circuit, we opt for ($\log N{=}16$, $\log q{=}1,440$). We focus on a fully packed ciphertext that comprises 512 database entries. The client needs $0.9$s to encode and encrypt the search query, and the server requires $30$s to run it on the encrypted database. The decryption/decoding executes at $140$ms, and the client in/out communication is $25$MB per ciphertext.

\descr{Results.} REP introduces a significant computational and communication overhead for the offloading client as the extended vector is encrypted into $\lambda$ ciphertexts. PE, on the contrary, less than triples the encoding/encryption time and communication overhead towards the server. The server computational overhead is $51\times$ more important for REP ($1,500$s), compared to the baseline. As the circuit is of depth $7$ (with relinearization), the convolutions required by the PE create significant computation and egress communication overheads for the server (\ie $83\times$ or $2,500$s) for computations and $131\times$ more communication than the baseline). Accordingly, PE's verifying cost is dominated by the numerous ciphertexts that store the encoding ($21$s). We observe that ReQ significantly reduces by more than $20$-fold the server and client computational overhead while minimizing the communication overhead, compared to the standard PE approach (at the cost of online client-server interactions). With ReQ, the server evaluation now takes only $120$s with a short client involvement during the interaction ($4.9$s). The client verification eventually requires only $550$ms ($5.5$s in total compared to $21$s for standard PE). REP does not need this optimization but needs more challenge values to be verified in the \textbf{Verify} phase.

\subsubsection{Machine-Learning Prediction}\label{sec:eval:inference}

Advances in machine learning (ML) enable new ways for data analytics and several works explore the use of HE in order to protect, during ML inference, data confidentiality~\cite{brutzkus2019low,madi2020computing,niu2020toward,xu2020secure,zhang2020privacy}. However, as the correctness of the prediction can be tantamount to confidentiality, we apply \name to an encrypted ML inference pipeline. Indeed, misclassification or malicious predictions could render the application pointless and have dire consequences, such as financial misprediction or cyberthreat misclassification, for the end users~\cite{ghodsi2017safetynets,madi2020computing,weng2021mystique,xu2020secure}. 

\descr{Description.} 
For this use-case, we re-implemented the Low-Latency Privacy Preserving Inference (LoLa)~\cite{brutzkus2019low} pipeline in Go. This pipeline uses a neural network composed of a convolutional layer (with 5 kernel maps of size $5{\times}5$ and a stride of 4), followed by a square activation function and a fully connected layer. The model is pre-trained on the MNIST hand-written digit dataset (comprising $28 {\times} 28$ grayscale images) and its weights are offloaded encrypted by an ML provider. Following~\cite{brutzkus2019low}, the client encodes and packs each image into 25 ciphertexts with different packing approaches that facilitate the neural-network operations. 

\descr{Parameterization.} 
We use the HE parameters ($\log N{=}15$, $\log q{=}700$). The client requires $1.6$s to encode and encrypt the input image, and the server requires $13$s to perform the inference. The decryption/ decoding executes at $330$ms and the server ingress/egress communication is $160$/$63$MB. We set $\lambda{=}64$ for REP and $\log t{=}56$ for PE. 

\descr{Results.} 
Figure~\ref{fig:lola} shows that, as the ciphertexts are not fully packed, REP introduces a negligible overhead for the offloading client (less than $1\times$). The PE creation is $2\times$ slower than the baseline. Whereas REP is seamless for the server, PE introduces an overhead that is $3\times$ more; the activation function requires a multiplication that introduces convolutions. The verification cost for REP is piloted by the challenge verification and the decoding, which leads to a total \textbf{Verify} phase of $0.6$s. PE verifies the challenges in $0.06$s but introduces decryption and decoding overhead due to the exchange of $3\times$ more ciphertexts than the baseline. We observe that PoC reduces the communication to only two ciphertexts assuming an investment from the server (\ie $8\times$ slower evaluation). 

\subsubsection{Federated Learning}\label{sec:eval:fl} 

Federated learning has emerged as a technique for improving ML training by relying on silos of training-data distributed amongst several clients. Clients locally train models and exchange the model updates with a central entity that aggregates them and broadcasts the result. As the values exchanged during the training of a federated model leak information about the training data~\cite{boenisch2021curious,melis2019exploiting,nasr2019comprehensive,zhu2020deep}, several works rely on secure aggregation techniques by using HE to protect confidentiality~\cite{fereidooni2021safelearn,zheng2019helen,zhang2020batchcrypt}. Also, ensuring the correctness of the aggregation is crucial for avoiding the introduction of backdoors, maliciously biased weights, and/or simply the deliberate exclusion of some client models by a malicious server~\cite{boenisch2021curious,pasquini2022eluding,tramer2022truth,xu2019verifynet,zhang2020privacy}.

\descr{Description.} 
We adapt the federated averaging pipeline (FedAvg) proposed by McMahan \etal~\cite{mcmahan2017communication} for MNIST digit recognition to a cross-silo setting where clients trust each other but not the aggregator. This pipeline employs a simple multi-layer perceptron with two hidden-layers of 200 units using the ReLu activation function ($199,210$ parameters in total). The dataset is split across 100 clients, and 10 of them are randomly polled at each epoch for the federated averaging. Weights are locally quantized, by each client, to integers (16-bits as in~\cite{zhang2020batchcrypt}) before being sent encrypted to the aggregator. After secure aggregation, the global weights are then decrypted and de-quantized by the clients and used for the next epoch. After 100 epochs, we obtain an accuracy of $91\%$ (achieving higher accuracy with more iterations or different parameterization is out of the scope of this work). As all the clients trust each other, one client is in charge of generating the cryptographic keys and uses a secure channel (such as TLS) to share them via the server with the other clients. As the local training datasets are never shared, \name enables result verification without needing the input data.

\descr{Parameterization.} 
We use the HE parameters ($\log N{=}15$, $\log q{=}700$) and report results per epoch. Clients require $4.3$s to encode and encrypt their local models, and the server requires $0.2$s to aggregate them. The client decryption/decoding requires $214$ms and the client egress/ingress communication is $44$/$44$MB. We set the authenticator parameters to $\lambda{=}64$ for REP and $\log t{=}55$ for PE. 

\descr{Results.} 
Figure~\ref{fig:FL} shows that, due to the huge volume of data ($199,210$ model parameters encrypted in $7$ ciphertexts per client), REP creation is about $60\times$ slower than the baseline, thus making the PE encoding much more efficient ($255$s vs. $11$s, resp.). Similarly, the non-existent multiplicative complexity of secure aggregation showcases that PE is more efficient for the server; it introduces a $1.5\times$ server evaluation overhead, compared to REP's $80\times$ one ($0.4$s vs. $13$s). Accordingly, the client verification with PE is only $2.2\times$ slower than the baseline, whereas REP introduces an overhead of $70\times$ more ($0.7$s vs. $15$s resp.). The overall communication overhead of REP is more significant than PE ($2.4$GB vs. $88$MB).

\subsection{Comparison With Prior Work}\label{sec:eval:prior}

A direct performance comparison between \name and the related work is not straightforward due to (i)~the different models employed for verifying the integrity of homomorphic computations, (ii)~the related work's limitations regarding the homomorphic operations supported (see Table~\ref{tab:rw}), and (iii)~the lack of implemented solutions. For instance, Fiore \etal~\cite{fiore2020boosting} cannot support the HE parameterization used in \S\ref{sec:eval} and Bois \etal~\cite{bois2021flexible} cannot evaluate rotation or relinearization operations; neither work has a public implementation. 
We implemented Rinocchio~\cite{ganesh2021rinocchio} and discovered that it can practically handle only arithmetic operations: To support the rounding operations required by modern HE schemes for tensoring and rotations (key-switching) (see \S\ref{sec:prelim:fhe}), Rinocchio requires emulating multiple rings which would significantly slow down its performance. Thus, we evaluate Rinocchio only on the federated learning use-case (\S\ref{sec:eval:fl}). Although it yields a succinct proof ($118$kB), Rinocchio is computationally more expensive than our solution ($23$s and $8$s for the Create and Verify phases resp. compared to $0.4$s and $0.7$s for \namecomma) -- and assuming the already expensive setup phase ($\sim$1h). Only the early work of Fiore \etal~\cite{fiore2014efficiently} has been officially implemented. As their work is limited to quadratic functions and cannot support rotations, we can compare it with \name only on the ride-hailing (\S\ref{sec:eval:rh}) and federated learning (\S\ref{sec:eval:fl}) use-cases. For a similar HE parameterization, we estimate that their verification would take $180$ms for the ride-hailing (resp. $4.1$s for federated learning), whereas \name requires $50$ms with the best encoding (resp. $0.7$s for federated learning). Similarly, following Fiore \etal's evaluation, the server's ride-hailing evaluation takes $0.8$s (resp. $0.1$s for federated learning), which is very similar to \name ($0.5$s and $0.4$s, resp.). We stress again that none of the other use-cases from  \S\ref{sec:eval} can be efficiently achieved by using prior work.

\subsection{Evaluation Take-Aways}\label{sec:eval:summ}
Our evaluation shows that \name is suitable for various applications and yields different trade-offs depending on the setting and the security requirements. When the volume of input data is small and REP's (\S\ref{sec:vche1}) extended vector fits in a single ciphertext, then REP outperforms PE (\S\ref{sec:eval:rh},\ref{sec:eval:inference}). When the application requires fully packed ciphertexts, PE (\S\ref{sec:vche2}) reduces the load of the client by an order of magnitude (\S\ref{sec:eval:search},\ref{sec:eval:fl}). Furthermore, our analysis demonstrates that PE yields a significant overhead for the server when the circuit multiplicative depth is large (\S\ref{sec:eval:search}). This is alleviated by ReQ (\S\ref{sec:vche2:RQ}) which further improves the client verification time when interactivity is possible. With some computation overhead at the server, the PoC (\S\ref{sec:vche2:PP}) reduces the volume of data sent back to the decrypting client (\S\ref{sec:eval:inference}). Both ReQ and PoC optimizations improve the computation and communication overheads at the client, thus making the use of \name suitable for constrained clients. Our evaluation also shows that \name can cope with various use-cases relying on complex homomorphic operations that were not supported by the state of the art~\cite{bois2021flexible,fiore2020boosting,ganesh2021rinocchio}. Overall, we observe that it enables homomorphic computation verifiability with acceptable overheads for the client and server (and even more considering the overhead of the HE pipeline itself; see Appendix~\ref{ap:disc:perspective}). For instance, it enables the verifiability of a disease prediction result on genomic data, with less than $3\times$ computation and communication overhead for the client and the server, compared to the HE baseline.

\section{Related Work}\label{sec:rw}
We review the literature on the verification of HE operations. We identify the following approaches.

\descr{Enc-and-Mac~\cite{lai2014verifiable,li2018privacy,li2021toward}.} These works create homomorphic authenticators of the plaintext data and its encryption concurrently. To preserve confidentiality, the homomorphic authenticator needs to provide semantic security. Li \etal's solutions~\cite{li2018privacy,li2021toward} work only for bounded degree polynomials and cannot cope with non-algebraic HE operations, \eg relinearization or key-switching, and are as secure as the underlying MAC.

\descr{Enc-then-Mac~\cite{catalano2014authenticating,tran2016efficient,xu2017cryptanalysis,fiore2014efficiently}.} Catalano \etal~\cite{catalano2014authenticating} authenticate linear ciphertext operations using the homomorphic MAC of~\cite{catalano2013practical}. By modifying this authenticator and by relying on different security assumptions (\eg structure-preserving signatures~\cite{libert2013linearly}, error-free approximate greatest common divisor problem~\cite{joo2014homomorphic}, and functional encryption~\cite{goldwasser2013run}), other works support quadratic functions~\cite{tran2016efficient,xu2017cryptanalysis}. Fiore \etal~\cite{fiore2014efficiently} propose the first solution tailored to a RLWE-based scheme similar to BGV: Ciphertexts are integrity-protected via a homomorphic MAC~\cite{backes2013verifiable} and a hashing technique compresses them to save storage and computational resources. But their instantiation works for a simplified version of BV without relinearization and key-switching (thus, without rotations nor SIMD) and supports only quadratic functions (see Table~\ref{tab:rw}).

\descr{SNARKs~\cite{fiore2020boosting,bois2021flexible,ganesh2021rinocchio}.} These works consider proving the correctness of ciphertext operations by relying on generic proof systems \eg~\cite{goldwasser2015delegating,ben2014succinct,bunz2018bulletproofs,chiesa2020fractal,groth2018updatable,maller2019sonic,parno2013pinocchio,setty2020spartan,wahby2018doubly,weng2021wolverine}. Fiore \etal\cite{fiore2020boosting} use commitments and SNARKs to achieve verification of homomorphic evaluation for bounded polynomial operations. They reduce the overhead due to the ciphertext expansion by relying on two blocks: (i)~a commit-and-prove SNARK for arithmetic circuits, and (ii)~a commit-and-prove SNARK for multiple polynomial evaluations. But their solution does not support modulus switches and rounding operations; contrary to our approach, this limits the use of several state-of-the-art schemes (\eg BFV and BGV). Moreover, it requires the SNARKs and the HE scheme to share the same cryptographic parameters, which can result in an undesired overhead; the HE modulus should match the SNARK one which is usually much larger. Although this latter constraint is lifted by Bois \etal\cite{bois2021flexible}, the resulting approach still limits the admissible HE pipelines (since it does not support modular reduction). Rinocchio~\cite{ganesh2021rinocchio} introduces a new SNARK for rings that can be used for verifiable computations on encrypted data. Compared to~\cite{fiore2020boosting}, Rinocchio supports common lattices used by HE schemes. But Rinocchio is still not as expressive as HE and cannot practically support core operations (\eg relinearization, key-switch, and bootstrapping). Overall, these works are limited in the admissible operations supported (see Table~\ref{tab:rw}) and potentially induce a non-negligible memory and computation cost for the prover.

\descr{Trusted Hardware~\cite{natarajan2021chex}.} This line of research employs trusted hardware to ensure computation integrity. But it relies on a different threat model (\ie a trusted chip manufacturer) and might be vulnerable to side-channel privacy leakage~\cite{Lipp2021Platypus,wang2017leaky,xu2015controlled} and integrity attacks~\cite{fei2021security,randmets2021overview}.

\section{Conclusion}\label{sec:conc}
In this work, we have proposed two error-detection encodings that can be used to verify the integrity of homomorphic computations. With these encodings, a client can augment existing privacy-preserving pipelines based on homomorphic encryption with computation verification at a minimal cost; this way, their threat model can be easily extended from an honest-but-curious adversary provider to a malicious but rational one. We have also provided \namecomma, a new open-source library that implements our solution and we demonstrated its practicality and versatility on several uses-cases.

\section*{Acknowledgements}\label{veritas:ack}
This work was supported in part by grant 200021\_178978/1 (PrivateLife) of the Swiss National Science Foundation (SNF). 

\bibliographystyle{IEEEtranS}
\bibliography{literature}

\begin{appendices}
\section{Homomorphic Authenticators}\label{ap:HA}
In this section, we recall the formalism of homomorphic authenticators. 

\descr{$\text{HA.KeyGen}$}$(1^\lambda)\rightarrow (\text{evk}, \text{sk})$: Output a secret key $\text{sk}$ and an evaluation key $\text{evk}$ for the authenticator. 

\descr{$\text{HA.Auth}$}$(m, \tau; \text{sk})\!\rightarrow\! \sigma$: Create an error-detecting authentication $\sigma$ for a message $m$ associated with an identifier~$\tau$. 

\descr{$\text{HA.Eval}$}$(f(\cdot), \vv{\boldsymbol{\sigma}};\text{evk})\rightarrow \sigma'$: Evaluate a deterministic function $f(\cdot)$ on authenticated data where $\vv{\boldsymbol{\sigma}}{=}(\sigma_1, \!..., \sigma_n)$ and each $\sigma_i$ authenticates the $i$-th input $m_i$. $\sigma'$ authenticates the result  $m'{=}f(m_1, \!..., m_n)$. 

\descr{$\text{HA.Ver}$}$(\mathcal{P}, \sigma'; \text{sk}){\rightarrow}\{0,1\}$: Check that $m'$ obtained from the authentication $\sigma'$ is the correct output of the program $\mathcal{P}{=}(f(\cdot), (\tau_1, \!... , \tau_n))$; \ie the evaluation of $f(\cdot)$ on authenticated inputs identified by $\tau_1, \!... , \tau_n$.

\begin{figure}
\refstepcounter{expe}\label{expe}
    \experiment{$\textbf{Exp}_\mathcal{A}[\text{HA}, \lambda]$}{
    ~
    \begin{enumerate}[leftmargin=*]
        \item $\textbf{HA.KeyGen}(1^\lambda)$: The challenger generates the keys $(\text{evk}$ and $\text{sk})$ and initializes the list of authenticated inputs: $T{=}\emptyset$.
        \item $\textbf{HA.Auth}(m, \tau; \text{sk})$: $\mathcal{A}$ asks the challenger for authentication of $m$ with identifier $\tau$. If the identifier was already queried (\ie $(\tau, \cdot){\in}T$), then the challenger aborts. Otherwise, it returns $\sigma\leftarrow\textbf{HA.Auth}(m, \tau; \text{sk})$ and appends $(m, \tau)$ to $T$.
        \item $\textbf{Forgery}$: The adversary $\mathcal{A}$ outputs a forgery  $(\mathbf{m}^*,\mathcal{P}^*=(f,(\tau_1^*, \!..., \tau_{n}^*), \boldsymbol{\sigma}^*)$.
        The experiment outputs $1$ if the verification accepts and that either (i)~the program $\mathcal{P}^*$ is not well-defined on $T$ or that (ii)~the output $\mathbf{m}^*$ does not match the correct output of $\mathcal{P}^*$ evaluated on the authenticated data in $T$.
        \end{enumerate}
    }
\end{figure}

\noindent We now recall some of the properties of an HA scheme. 
\begin{itemize}[leftmargin=*]
    \item \descr{Authentication Correctness:} For any message $m$ associated with $\tau$,
    
    \begingroup
    \[\Pr\!\left[ \text{HA.\!Ver}(\mathcal{I}_\tau\!, \boldsymbol{\sigma}\!; \text{sk}) {=} 1 \middle\vert\!\! \begin{array}{ll}
                  \!(\text{evk}, \text{sk})\!\leftarrow\! \text{HA.KeyGen}(1^\lambda)\\
                  \boldsymbol{\sigma} \leftarrow\text{HA.Auth}(m,\tau; \text{sk})\\
                \end{array}\!\!\!
              \right] \!\!{=} 1,\]
    \endgroup
    with {\small{$\mathcal{I}_\tau{=}(Id, \tau)$}} the program associated with the identity function.
    
    \item \descr{Evaluation Correctness:} Consider any key pair $(\text{evk}, \text{sk})$ generated through the $\text{HA.KeyGen}(1^\lambda)$ procedure. Define any fixed circuit $f(\cdot)$, and any correctly generated triplet $\{(\mathcal{P}_i, m_i, \sigma_i)\}_{i=1}^n$. If 
    $m^* {:=} f(m_1,\!...,m_n)$, $\mathcal{P}^* {:=}f(\mathcal{P}_1, \!..., \mathcal{P}_n)$, and
    $$\sigma^* :=\allowbreak \text{HA.Eval}(f,(\sigma_1, \!...,\sigma_n); \text{evk}),$$
    then $\text{HA.\!Ver}(\mathcal{P}^*, \sigma^*; \text{sk}){=}1$.

    \item \descr{Authenticator Security:} Given the security parameter $\lambda$, the probability of a malicious adversary convincing the verifier to accept a wrongfully computed result is negligible. 
    More formally, define {\small{$\textbf{Exp}_\mathcal{A}[\text{HA}, \lambda]$}} as in Experiment~\ref{expe}. The authenticator $\text{HA}$ is said to be \textit{secure} if for any PPT adversary $\mathcal{A}$, {\small{$\Pr[\textbf{Exp}_\mathcal{A}[\text{HA}, \lambda]\!=\!1] \!\leq\! \text{negl}(\lambda)$}}.
\end{itemize}

\section{Security Proof for REP (\S\ref{sec:vche1:def})}\label{ap:vche1}
\descr{Theorem~\ref{th:REP}:} Let $\lambda$ be a power-of-two security parameter. If the PRF $F_K$ and the canonical HE scheme are at least $\lambda$-bit secure, then for any admissible program $\mathcal{P}$, $\text{REP}$ as in Scheme~\ref{scheme:REP} is a secure authenticator and a PPT adversary has a probability of successfully cheating the verification negligible in $\lambda$. 

\begin{proof}
We follow Gennaro and Wichs' Theorem 3.1 on the security of their homomorphic MAC~\cite{gennaro2013fully}.  
Let $\mathcal{A}(1^\lambda)$ be a probabilistic polynomial time (PPT) attacker. We define the following game following Experiment~\ref{expe}: 

\descr{Game0:} This game is the forgery game based on Experiment~\ref{expe} as $\textbf{Exp}_\mathcal{A}[\text{REP}, \lambda]$. We recall that the game outputs $1$ if the verification procedure $\text{REP.Ver}(\mathcal{P}^*, \boldsymbol{\sigma}^*; \mathbf{sk})=1$ and one of the two conditions holds:
\begin{itemize}[leftmargin=*]
    \item \textbf{\textsf{Type~1}:} $\mathcal{P}^*$ is not \textit{well-defined} w.r.t. the set of inputs $T$ (see \S\ref{sec:prelim:verprog}).
    \item \textbf{\textsf{Type~2}:} $\mathcal{P}^*$ is \textit{well-defined} on $T$ and $f(\mathbf{m}_1^*, \!..., \mathbf{m}_n^*) {\neq} \mathbf{m}^*$ (\ie wrongful computation).
\end{itemize}
\noindent The scheme $\text{REP}$ is said secure if for all $\text{PPT} \,\mathcal{A}$, 
\begingroup
\[\Pr\left[ \text{Game0}(1^\lambda){=}1\right]{\leqslant} \text{negl}(\lambda).\]
\endgroup
Now let us design hybrid games by modifying Game0.

\descr{Game1:} We modify Game0 by replacing the PRF $F_K$ with random values in the authentication and verification procedures such that a random oracle outputs random values on the fly. By the pseudorandomness of $F_K$, it is clear that 
\begingroup
\[\Pr\left[\text{Game1}(1^\lambda)=1\right] \geqslant \Pr\left[\text{Game0}(1^\lambda)=1\right] - \text{negl}(\lambda).\]
\endgroup

\descr{Game2:} We change the winning condition of Game1 such that the adversary wins only if the \textsf{Type~2} attacks succeed and \textsf{Type~1} fail. 
Let $E$ be the event of winning a \textsf{Type~1} forgery in Game1. 
For the adversary to win a \textsf{Type~1} forgery, it must pass some unauthenticated inputs. Because the verifier can directly reject inputs with unknown identifiers, the only option for the adversary is to submit a forgery of the input authentications. Forging an input authentication implies: (i)~finding all the $\lambda/2$ replication slots to store the forged input and (ii)~guessing the $\lambda/2$ challenge values in $\mathbb{Z}_t$. The probability of (i)~is negligible by construction, and the probability of (ii)~is at most $(1/t)^{\lambda/2}$, which is negligible in $\lambda$ for any non-trivial plaintext modulus $t>3$.  
Therefore, we have that $\Pr\left[\text{Game2}(1^\lambda)=1\right] \geqslant \Pr\left[\text{Game1}(1^\lambda)=1\right] - \text{negl}(\lambda)$.

\descr{Game3:} We modify the winning condition of Game2. Now, the challenger remembers the authentication $\boldsymbol{\sigma}$ associated to a message $\mathbf{m}$ with identifier $\boldsymbol{\tau}$ (\ie it stores $(\boldsymbol{\tau}, \mathbf{m}, \boldsymbol{\sigma}){\in} T$ and remembers the randomness used for the encryptions and the keys). 
Additionally, the challenger re-packs the extended vectors such that $\forall i \in [\lambda]\setminus S $, $\mathbf{p}_i$ either stores the original message $\mathbf{m}$ or $\forall i \in S $, $\mathbf{p}_i$ stores a challenge vector (this re-packing is a client-side operation and does not change the adversary's view). 
Consider the adversary outputting a forgery 
$(\mathbf{m}^*,\mathcal{P}^*\!{=}(f(\cdot),(\boldsymbol{\tau}_1^*, \!..., \boldsymbol{\tau}_{n}^*)), \boldsymbol{\sigma}^*\!{=}\mathbf{c}^*)$. 
Denote by $\hat{\mathbf{c}}$ the honest ciphertext for the labeled program $\mathcal{P}^*$ (\ie the homomorphic evaluation of $f(\cdot)$ on inputs $\mathbf{c}_1, \!...,\mathbf{c}_n $). Denote by $\mathbf{p}^*_1$ the forgery of the first polynomial decrypted from $\mathbf{c}^*$. We modify the verification procedure to: 
\begin{enumerate}[leftmargin=*] 
    \item Use the stored ciphertexts in $T$ and decrypt them to obtain the challenge values, compute $f(\cdot)$ to get the challenge values of the output $\tilde{r}_{i}$, and recompute the result ciphertext $\hat{\mathbf{c}}$. Check if $\forall i {\in} S$, $\mathbf{p}^*_i{=}\tilde{\mathbf{p}}_{i}$ and the ciphertexts match. Otherwise, reject.
    \item $\forall i{\in} [\lambda]\setminus S$, check if all $\mathbf{p}^*_i$ are equal to the honest output obtained from $\hat{\mathbf{c}}$ (say $\mathbf{m}$). If so, reject.  
\end{enumerate}
By construction, any \textsf{Type~2} forgery accepted in Game2 is also accepted in Game3. This is because the protocol admits as functions only admissible polynomial operations on the plaintext space (the randomness of the ciphertexts is now fixed in this game) and aborts if the result is constant over the choice of the challenges. 
Thus, $\Pr\left[\text{Game3}(1^\lambda)=1\right] \geqslant \Pr\left[\text{Game2}(1^\lambda)=1 \right]$ and 
    \begingroup
\[\Pr\left[\text{Game3}(1^\lambda)=1\right] \geqslant \Pr\left[\text{Game0}(1^\lambda)=1 \right]-\text{negl}(\lambda).\]
\endgroup

\descr{Game4:} We modify Game3 such that, when answering the authentication queries, the challenger encrypts the actual message regardless of the slot index (\ie regardless of the set $S$). By the semantic security of the homomorphic encryption scheme, the adversary cannot distinguish if a specific ciphertext encrypts $\mathbf{m}$ or a challenge vector. 
Given a plaintext that either stores the original slot message or a challenge value, we can embed it into the authentication procedure for each position $i{\in} S$ and simulate either Game3 or Game4.
Suppose $\Pr\left[\text{Game4}(1^\lambda){=}1\right] < \Pr\left[\text{Game3}(1^\lambda)=1\right] {-}\text{negl}(\lambda)$. This would lead to $$\text{negl}(\lambda) \leqslant |\Pr\left[\text{Game3}(1^\lambda)=1\right]{-}\Pr\left[\text{Game4}(1^\lambda)=1\right]\!|$$ thus breaking the semantic security of the HE scheme. 

In Game4, $S$ is never used at authentication time. Thus, we can think of the challenger picking $S$ only at verification time. 
For any \textsf{Type~2} forgery $(\mathbf{m}^*,\mathcal{P}^*{=}(f,(\boldsymbol{\tau}_1^*, \!..., \boldsymbol{\tau}_{n}^*)), \boldsymbol{\sigma}^*)$, decryption leads to $\text{HE.Dec}(\mathbf{c}^*; \mathbf{sk}_{\text{HE}}) {=} (\mathbf{p}^*_1, \dots, \mathbf{p}^*_{\lambda})$.
Denote by $\hat{\mathbf{p}}_i$ the $i$-th  vector of the honestly generated ciphertext.
Let $S'{=} \{i \in[\lambda]: \mathbf{p}_i^* {=} \hat{\mathbf{p}}_i \wedge \mathbf{c}_i^* {=} \hat{\mathbf{c}}_i\}$ be the indices on which the forged and honest extended vectors match.
The adversary wins if the second and third steps of Game3 pass which only occurs if $S=S'$ (the protocol aborts if all the challenges have the same value). Since the program is admissible, this event happens with probability $2^{-\lambda}$. Thus, $\Pr\left[\text{Game4}(1^\lambda)=1\right] \leqslant 2^{-\lambda}$. In turn, we see that 
    \begingroup
 \[\Pr\left[\text{Game0}(1^\lambda)=1\right] \leqslant 2^{-\lambda}+ \text{negl}(\lambda) \leqslant \text{negl}(\lambda)\]
 \endgroup
 which can be parameterized to be negligible in $\lambda$, concluding the proof.
 
\end{proof}

\section{Security Proof for PE (\S\ref{sec:vche2:def})}\label{ap:vche2}
\descr{Theorem~\ref{th:pe}:} If the PRF $F_K$ and the canonical HE scheme are at least $\lambda$-bit secure and if $t$ is a $\lambda$-bit prime number, then, for any program $\mathcal{P}$ with authentications of bounded degree, $\text{PE}$ is a secure authenticator and a PPT adversary has a probability of successfully cheating the verification negligible in $\lambda$.

\begin{proof}
We follow the security analysis sketched in Catalano and Fiore's information-theoretic homomorphic MAC~\cite{catalano2013practical}. 
We consider a PPT adversary $\mathcal{A}(1^\lambda)$. We design the following series of games.

\descr{Game0:} This game is the forgery game based on Experiment~\ref{expe} as $\textbf{Exp}_\mathcal{A}[\text{PE}, \lambda]$. The different types of forgeries (\ie \textsf{Type~1} and \textsf{2}) are defined in Appendix~\ref{ap:vche1}.
Recall that the scheme $\text{PE}$ is said secure if $\forall \text{ PPT } \mathcal{A}$, 
\begingroup
\[\Pr\left[ \text{Game0}(1^\lambda)=1\right]{\leqslant} \text{negl}(\lambda).\]
\endgroup
Now let us design hybrid games by modifying Game0.

\descr{Game1:} We modify the verification procedure in Game0 such that the challenger checks whether the program $\mathcal{P}$ is well-formed or not (\ie all inputs are authenticated) using probabilistic polynomial identity testing (see~\cite{cryptoeprint:2015:194} Prop.1):
\begingroup
\[\left|\Pr[\text{Game1}(1^\lambda)=1]-\Pr[\text{Game0}(1^\lambda)=1]\right| \leqslant 2^{-\lambda}.\]
\endgroup
 
\descr{Game2:} We modify Game1 by replacing the PRF $F_K$ with random values in the authentication and verification procedures such that a random function outputs random values on the fly. By the pseudorandomness of $F_K$, it is clear that 
\begingroup
\[\left|\Pr\left[\text{Game2}(1^\lambda)=1\right] - \Pr\left[\text{Game1}(1^\lambda)=1\right]\right| \leqslant \text{negl}(\lambda).\]
\endgroup

\descr{Game3:} We modify only the verification procedure of Game2. For the labeled program $\mathcal{P}=(f(\cdot), (\boldsymbol{\tau}_1, \!..., \boldsymbol{\tau}_n))$, and a verification query $(\mathbf{m}, \mathcal{P}, \boldsymbol{\sigma}')$, parse $\boldsymbol{\sigma}'=(\mathbf{c}_0, \!..., \mathbf{c}_d)$. 
\begin{itemize}[leftmargin=*]
    \item If $\mathbf{y}_0=\text{HE.Dec}(\mathbf{c}_0;\mathbf{sk}_{\text{HE}}) \neq \mathbf{m}$, then reject.
    \item If $\mathcal{P}$ is not well-defined on $T$, then the challenger sequentially:
\begin{enumerate}
    \item $\forall i \in [n]$, s.t. $(\boldsymbol{\tau}_i,\cdot)\notin T$, samples uniformly at random 
    $\mathbf{r}_{\boldsymbol{\tau}_i}$ and computes $\boldsymbol{\rho}=f(\mathbf{r}_{\boldsymbol{\tau}_1}, \!..., \mathbf{r}_{\boldsymbol{\tau}_{n}})$.
    \item Decrypts the ciphertexts s.t. $\forall i \!\in\! [0{:}d]$, $\mathbf{y}_i = \text{HE.Dec}(\mathbf{c}_i; \mathbf{sk}_{\text{HE}})$.
    \item Computes $\mathbf{Z} {=} \boldsymbol{\rho} - \sum_{i=0}^{d} \mathbf{y}_i \cdot \alpha^i$.
    \item If $\mathbf{Z}{=}\mathbf{0}\!\!\mod t$ then the challenger accepts (\ie outputs 1), otherwise it rejects.
\end{enumerate}
\end{itemize}
As it is merely a syntactic change from Game2,
\begingroup
\[\Pr[\text{Game2}(1^\lambda)=1]\equiv \Pr[\text{Game3}(1^\lambda)=1].\]
\endgroup

\descr{Game4:} We modify the verification procedure of Game3. Consider that $\mathcal{P}$ is well-defined over $T$. For all $i\in [n]$ such that $(\boldsymbol{\tau}_i, \cdot)\notin T$ (\ie one of the slot values was not in $T$), the challenger chooses a dummy $\boldsymbol{\sigma}_i$ generated for a random message for the corresponding slot. The challenger then computes $\boldsymbol{\hat{\sigma}}'=(\mathbf{\hat{c}}_0, \!..., \mathbf{\hat{c}}_d)\leftarrow \text{PE.Eval}(f(\cdot), \vv{\boldsymbol{\sigma}}; \mathbf{evk})$. Now: 
\begin{itemize}[leftmargin=*]
    \item \raisebox{.5pt}{\textcircled{\raisebox{-.6pt} {\small A}}}~If $\forall k {\in} [0{:}d]$ $\text{HE.Dec}(\mathbf{c}_k;\mathbf{sk}_{\text{HE}}){=}\text{HE.Dec}(\mathbf{\hat{c}}_k;\mathbf{sk}_{\text{HE}})$, then accept. 
    \item \raisebox{.5pt}{\textcircled{\raisebox{-.6pt} {\small B}}}~If $\exists k$ s.t. $\text{HE.Dec}(\mathbf{c}_k;\mathbf{sk}_{\text{HE}}) {\neq} \text{HE.Dec}(\mathbf{\hat{c}}_k;\mathbf{sk}_{\text{HE}})$, compute $$\mathbf{Z} {=}\! \sum_{i=0}^{d}  \left(\text{HE.Dec}(\mathbf{c}_i;\mathbf{sk}_{\text{HE}}) {-} \text{HE.Dec}(\mathbf{\hat{c}}_i;\mathbf{sk}_{\text{HE}})\right)\!{\cdot} \alpha^i.$$
    If $\mathbf{Z}{=}\mathbf{0}\!\!\mod t$ then accept, otherwise reject. 
\end{itemize}
We show that the adversary has the same view as in Game3. Consider $(\mathbf{m},\mathcal{P}, \boldsymbol{\sigma}'=(\mathbf{c}_0, \!..., \mathbf{c}_d))$ with $\mathcal{P}=(f, (\boldsymbol{\tau}_1, \!..., \boldsymbol{\tau}_n))$ well-defined on $T$.
We distinguish two cases:

\begin{enumerate}[leftmargin=*]
    \item \textbf{All inputs were authenticated:} $\forall i \in [n]$, $(\boldsymbol{\tau}_i, \mathbf{m}_i)\in T$ and $\boldsymbol{\sigma}_i\leftarrow \text{PE.Auth}(\mathbf{m}_i, \boldsymbol{\tau}_i; \mathbf{sk})$. Recall that the challenger computes $\boldsymbol{\hat{\sigma}}'=(\mathbf{\hat{c}}_0, \!..., \mathbf{\hat{c}}_d)\leftarrow \text{PE.Eval}(f(\cdot), \vv{\boldsymbol{\sigma}}; \mathbf{evk})$. Now if \raisebox{.5pt}{\textcircled{\raisebox{-.6pt} {\small A}}} occurs, then the answer is correct by the correctness of the HE scheme. If \raisebox{.5pt}{\textcircled{\raisebox{-.6pt} {\small B}}} occurs, 
    as the same values $\{\mathbf{r}_{\boldsymbol{\tau}_i}\}$ are generated, $$\boldsymbol{\rho}=\text{PE.Ver}(\mathcal{P}, \boldsymbol{\hat{\sigma}}'; \mathbf{sk})=\text{PE.Ver}(\mathcal{P}, \boldsymbol{\sigma}'; \mathbf{sk}).$$ 
    So accepting if $\mathbf{Z}=\mathbf{0}$ is the same as returning the output of $\text{PE.Ver}(\mathcal{P}, \boldsymbol{\sigma}'; \mathbf{sk})$.
    
    \item \textbf{Some of the inputs were not authenticated:} $\exists i \in [n], $ s.t. $(\boldsymbol{\tau}_i,\cdot)\notin T$ (at least in one of its $N$ components). Thus, by definition of the well-defined program, wires with those inputs are not used in the computation (\ie same output regardless of the value of those wires). This also holds after the homomorphic transformation and thus the dummy input chosen does not impact the computation. 
\end{enumerate}
As Game4 is only a syntactic change from Game3, $\Pr[\text{Game3}(1^\lambda)=1]\equiv \Pr[\text{Game4}(1^\lambda)=1]$.

\descr{Game5:} We modify the verification procedure of Game4. We define a flag {\scriptsize{\ovalbox{\!\textsc{Bad}\!}}} initially set to false. When verifying $(\mathbf{m}, \mathcal{P}, \boldsymbol{\sigma}')$, if the computation $\mathbf{Z} {=} \mathbf{0}\!\!\mod t$, then the challenger rejects and sets {\scriptsize{\ovalbox{\!\textsc{Bad}\!}}}${=}$True. 
By definition of the verification procedure, \textsf{Type~1} and \textsf{Type~2} forgeries are rejected leading to $\Pr[\text{Game5}(1^\lambda)=1]{=}0$.
Note that Game4 and Game5 are identical unless the event $\xi{=}$\! ``~{\scriptsize{\ovalbox{\!\textsc{Bad}\!}}} is true'' occurs. Thus, 
$\Pr[\text{Game5}(1^\lambda){=}1 \wedge \neg \xi ] {=} \Pr[\text{Game4}(1^\lambda){=}1]$ leads to 
\begingroup
\[|\Pr[\text{Game4}(1^\lambda)=1]-\Pr[\text{Game5}(1^\lambda)=1]| \leqslant \Pr[\xi].\]
\endgroup

\noindent For the flag {\scriptsize{\ovalbox{\!\textsc{Bad}\!}}} to be set, the challenger needs to compute $\mathbf{Z}$. Depending on the definition of $\mathcal{P}$ two cases occur: 
\begin{enumerate}[leftmargin=*]
    \item $\mathcal{P}$ is well-defined and $\mathbf{Z}{=}\sum_{k=0}^{d}  (\mathbf{y}_k {-} \hat{\mathbf{y}}_k)\cdot \alpha^k{=}0\!\!\mod t$ where $\exists \hat{k}\in[0{:}d]$ s.t. $\mathbf{y}_{\hat{k}} \neq \hat{\mathbf{y}}_{\hat{k}}$. Call this constraint $\xi_1$.
    \item $\mathcal{P}$ is not well-defined, $\mathbf{Z}{=}\boldsymbol{\rho} - \sum_{k=0}^d \mathbf{y}_k \cdot \alpha^k{=}0\!\!\mod t$, and $\boldsymbol{\rho}$ is computed using at least one value $r_{\tau^*}$ that was not authenticated. Call this constraint $\xi_2$.
\end{enumerate}
This leads to 
\begin{equation}\label{eq:bad}
        \Pr[\xi] \leqslant  \Pr[\mathbf{Z}=0| \xi_1] + \Pr[\mathbf{Z}=0| \xi_2].
\end{equation}
Observe that before the verification, there exists exactly $t$ possible tuples $(\alpha, \{r_\tau\}_{\tau \in T})$ consistent with the adversary's view. 
The first probability in Eq.~\ref{eq:bad} is bounded by $d/t$ as the polynomial $\sum_{i=0}^d (\mathbf{y}_i-\hat{\mathbf{y}}_i) \cdot x^i$ has at most $d$ zeros and there are at most $t$ possible values for $\alpha$. Indeed, finding these values is equivalent to finding the common zeros to $N$ degree $d$ polynomial equations. By defining the bivariate polynomial $\mathcal{Z}[X,Y] {=} \sum_{k=0}^{d}((\mathbf{y}_k[Y] - \hat{\mathbf{y}}_k[Y])\cdot X^i)$, we can rewrite the above polynomial as $\mathcal{Z}[X,Y] {=} \sum_{i=0}^{d} \sum_{j=0}^N z_{ij}\cdot Y^jX^i$. To find the zeros in $X$ for all possible values in $Y$, we can project on the basis generated by the $\{Y^j\}_j$ leading to $N$ polynomial equations of the form $\sum_{i=0}^d z_{ij} x^i{=}0$, for $j {\in} [0{:}N{-}1]$, 
each with at most $d$ zeros. Thus, the overall number of zeros in the X-dimension of the bivariate polynomial is the intersection of all these zeros which cannot be more than $d$.
For the second probability of Eq.~\ref{eq:bad}, as the program is not well-defined, $\boldsymbol{\rho}$ can be seen as a non-constant polynomial in $\{r_{\tau^*}\}_{\tau \notin T}$. As no query has involved $\tau^*$, the adversary can only guess its value with probability $1/t$. By the polynomial identity lemma, $Pr[Z{=}0|\xi_2] {\leqslant} \frac{d}{t}$. 
Thus, $\Pr[\xi] \leqslant \frac{2d}{t}$, which in turn leads to
\begingroup
\[\Pr[\text{Game0}(1^\lambda)=1] \leqslant  \frac{2d}{t}{+}2^{-\lambda}{+}\text{negl}(\lambda) \leqslant \text{negl}(\lambda),\]
\endgroup
which can be parameterized to be negligible in $\lambda$, concluding the proof.
\end{proof}

\section{Security Proof for {PoC} (\S\ref{sec:vche2:PP})}\label{ap:PoC}
In this section, we analyse the security of our polynomial compression protocol (PoC).

\begin{theorem}\label{th:pp}
PE achieving the conditions of Th.\ref{th:pe} and combined with the polynomial compression of Fig.~\ref{proto:poly} is a secure authenticator~(\S\ref{sec:prelim:vche} and Appendix~\ref{ap:HA}).
\end{theorem}

\begin{proof}
We focus on the verification procedure of PE to estimate the probability of check (3) passing with wrongful polynomials. 
Let $\text{W}$ be the event that the verifier accepts.
Consider a malicious server that deviates from the honest prover. The view of the malicious prover (denoted by a tilde) is different from that of the honest one, \ie $\tilde{\boldsymbol{\sigma}} {\neq} \boldsymbol{\sigma}'$. Consider the messages sent by the server ($\textsf{m}_1$ and $\textsf{m}_2$) depend, respectively, on all the information previously sent by the client. For any message $k$, we define $\textit{E}_k$ as the event that the message $\tilde{\textsf{m}}_k$ sent by the prover agrees with the message $\textsf{m}_k$ that the honest prover would have sent on the same view $\tilde{\boldsymbol{\sigma}}$.
By the law of total probability, we can write $\text{Pr}[\text{W}] \leqslant \text{Pr}[\text{W}|\bar{\textit{E}_2}]+\text{Pr}[\text{W}|\textit{E}_2]$.
The probability $\text{Pr}[\text{W}|\bar{\textit{E}_2}]$ is bounded by 
\begingroup
\[\text{Pr}[\text{W}|\bar{\textit{E}_2}] \leqslant \text{Pr}\left[\boldsymbol{\rho}(\alpha){=}\sum_{i=0}^d \tilde{w}_i\cdot\alpha^i \middle\vert \tilde{\textsf{m}}_2\neq \textsf{m}_2, \alpha \leftarrow \mathbb{Z}_t^*\right].\]
\endgroup
Following the security proof of PE (Appendix~\ref{ap:vche2}), this probability is upper bounded by $d/(t{-}1)$ by the polynomial identity lemma.
The probability $\text{Pr}[\text{W}|\textit{E}_2]$ can be decomposed by the law of total probability as $\text{Pr}[\text{W}|\textit{E}_2] \leqslant \text{Pr}[\text{W}|\textit{E}_2\wedge\bar{\textit{E}_1}]+\text{Pr}[\text{W} | \textit{E}_2\wedge\textit{E}_1]$.

The probability $\text{Pr}[\text{W}|\textit{E}_2\wedge\bar{\textit{E}_1}]$ is bounded by
\begingroup
 \[\text{Pr}[\text{W}|\textit{E}_2\wedge\bar{\textit{E}_1}]{\leqslant}\text{Pr}\!\left[\!\!\!\begin{array}{c}\tilde{w}_0{=}\mathbf{m}(\delta)\\H_{\delta,\beta}(\boldsymbol{\sigma}')\stackrel{?}{=} \sum_{i=0}^d \tilde{w}_i {\cdot} \beta^i
\end{array}
\!\!\middle\vert\!\!\begin{array}{c}\tilde{\textsf{m}}_1{\neq} \textsf{m}_1\\ \tilde{\textsf{m}}_2{=} \textsf{m}_2\end{array}\!\!\!\!\right]\!\!{\leqslant} \frac{d{+}N}{t}.\]
\endgroup
The inequality holds by the property of the hash (which is itself derived from the polynomial identity lemma); the malicious prover's knowledge of $(\delta, \beta)$ arrives after $\tilde{\textsf{m}}_1$ (see~\cite[Th.2]{fiore2014efficiently} for the analysis of such polynomial hash collision probability). The probability $\text{Pr}[\text{W}|\textit{E}_2\wedge\textit{E}_1]$ is upper bounded by $\text{Pr}[\text{W}|\textit{E}_1]$ which is, by the polynomial identity lemma on $H_{(\cdot, 0)(\boldsymbol{\tilde{\sigma}})}$, bounded by $\frac{d+N}{t}$.
Overall, the probability that the verifier accepts the protocol conversing with the malicious prover is bounded by 
\begingroup
\[\text{Pr}[\text{W}] \leqslant \frac{2(d+N)}{t}{+}\frac{d}{t{-}1}.\]
\endgroup
As a result, with an appropriate choice of parameters, the prover can only cheat the verifier with negligible probability.
\end{proof}

\section{Security Proof for ReQ (\S\ref{sec:vche2:RQ})}\label{ap:ReQ}
In this section, we describe in more depth our interactive re-quadratization protocol (ReQ) shown in Fig~\ref{proto:RQ} and analyse its security.

Figure~\ref{proto:RQ} presents the ReQ protocol in detail. At the $G$-th multiplicative gate, the server holds $\boldsymbol{\sigma}' {=} (\mathbf{c}_0, \mathbf{c}_1, \mathbf{c}_2, \mathbf{c}_3, \mathbf{c}_4)$. It interacts with the client to obtain $\bar{\boldsymbol{\sigma}} {=} (\mathbf{c}_0, \bar{\mathbf{c}}_1, \bar{\mathbf{c}}_2)$. The server sends the client the higher degree terms $\mathbf{c}_3$ and $\mathbf{c}_4$. The client decrypts them to $\mathbf{y}_3$ and $\mathbf{y}_4$ by using its secret key $\mathbf{sk}_{\text{HE}}$. 
For security reasons, ReQ introduces random blindings that need to be accounted for in further computations.
It samples uniformly two random numbers $\kappa_1, \kappa_2 {\leftarrow} \mathbb{Z}_t$ and two random polynomials $\mathbf{r}, \bar{\mathbf{r}} {\leftarrow} \mathcal{R}_t$. It sets $\bar{\mathbf{y}}_2=\alpha\kappa_1 \mathbf{y}_3 + \alpha^2\kappa_2 \mathbf{y}_4  + \mathbf{r}$.
We introduce the $\text{Shift}(\cdot)$ function that takes as input the sub-circuit $\mathcal{P}_G$ that leads to gate $G$, the authenticator secret key $\mathbf{sk}$, the identifiers $\boldsymbol{\tau}$, and the set $\Omega$ of previously used randomness (initially empty). This function returns the polynomial offset that was introduced by the addition of blindings in previous re-quadratizations. We call $\Delta_G$ the result of $\text{Shift}(\mathcal{P}_G, \boldsymbol{\tau}, \mathbf{sk}, \Omega)$.
It then evaluates $\bar{\mathbf{y}}_1=\bar{P}(\alpha) = \alpha^3 \mathbf{y}_4 {+} \alpha^2 \mathbf{y}_3 {-} \alpha \bar{\mathbf{y}}_2 {-} \Delta_G {+} \bar{\mathbf{r}}$. 
It encrypts $\bar{\mathbf{c}}_2 = \text{HE.Enc}(\bar{y}_2; \mathbf{pk}_{\text{HE}})$ and $\bar{\mathbf{c}}_1 = \text{HE.Enc}(\bar{y}_1; \mathbf{pk}_{\text{HE}})$ and sends them to the server. 
It also updates the list $\Omega$ with $(\mathbf{r}, \bar{\mathbf{r}})$.
The server appends them to the corresponding ciphertexts (\ie $\bar{\mathbf{c}}_i {\leftarrow} \mathbf{c}_i {+} \bar{\mathbf{c}}_i$, for $i{\in}\{1,2\}$). 
To remove the final offset, the verification procedure at the client is slightly modified. As a result, the client can pre-process (at least part) of the offsets and use them directly during the interactive protocol. Let us now analyze ReQ's security.

\begin{theorem}\label{th:rq}
PE achieving the conditions of Th.\ref{th:pe} and combined with the ReQ protocol from Fig.~\ref{proto:RQ} is a secure authenticator (\S\ref{sec:prelim:vche} and Appendix~\ref{ap:HA}).
\end{theorem}

\begin{proof}
The security of PE remains unchanged by the correctness of the ReQ that follows from its construction: evaluated on the secret point $\alpha$, $\bar{\boldsymbol{\sigma}}$ equals the same evaluation of the original degree-four $\boldsymbol{\sigma}'$ (up to a deterministic shift $\Delta_G$). The quantity $\Delta_G \!=\!\text{Shift}(\mathcal{P}_G, \boldsymbol{\tau}, \mathbf{sk}, \Omega)$ enables us to remove the offset introduced by the previous randomness. Hence, we subtract it to the term of degree one.
As $\kappa_1$ and $\kappa_2$ are uniformly random values in $\mathbb{Z}_t$, the quantity $\alpha\kappa_1 \mathbf{y}_3 {+} \alpha^2\kappa_2 \mathbf{y}_4$ is a random linear combination of $\mathbf{y}_3$ and $\mathbf{y}_4$. The addition of the random polynomial $\mathbf{r}$ ensures perfect secrecy, hence the result $\bar{\mathbf{y}}_2$, and even more its encryption, reveal nothing about $\alpha$: $\bar{\mathbf{y}}_2$ is indistinguishable from a random value in $\mathcal{R}_t$. The random polynomial $\bar{\mathbf{r}}$ acts again as a blinding value thus providing perfect secrecy of the polynomial $\bar{\mathbf{y}}_1$. The server knows only that $\bar{\mathbf{c}}_1$ and $\bar{\mathbf{c}}_2$ encrypt $\bar{\mathbf{y}}_1$ and $\bar{\mathbf{y}}_2$, respectively. Although the (bivariate) polynomial $X^4{\cdot} \mathbf{y}_4 {+} X^3{\cdot} \mathbf{y}_3 {+} X^2{\cdot}(\mathbf{y}_2 {-} \bar{\mathbf{y}}_2) {+} X(\mathbf{y}_1{-}\bar{\mathbf{y}}_1 {+} \bar{\mathbf{r}} {-} \Delta_G)$ admits $\alpha$ as one of its roots for all $N$ dimensions, the server cannot find the roots as it does not know its coefficients; the random polynomial $\bar{\mathbf{r}}$ is kept secret. Consequently, both ciphertexts $\mathbf{\bar{c}}_1$ and $\mathbf{\bar{c}}_2$ reveal nothing about the secrets (\ie $\mathbf{sk}, \Omega$).
\end{proof}

\section{Security Configuration vs. Overhead}\label{ap:overhead}

\begin{figure}
    \centering
    \includegraphics[width=.4\textwidth]{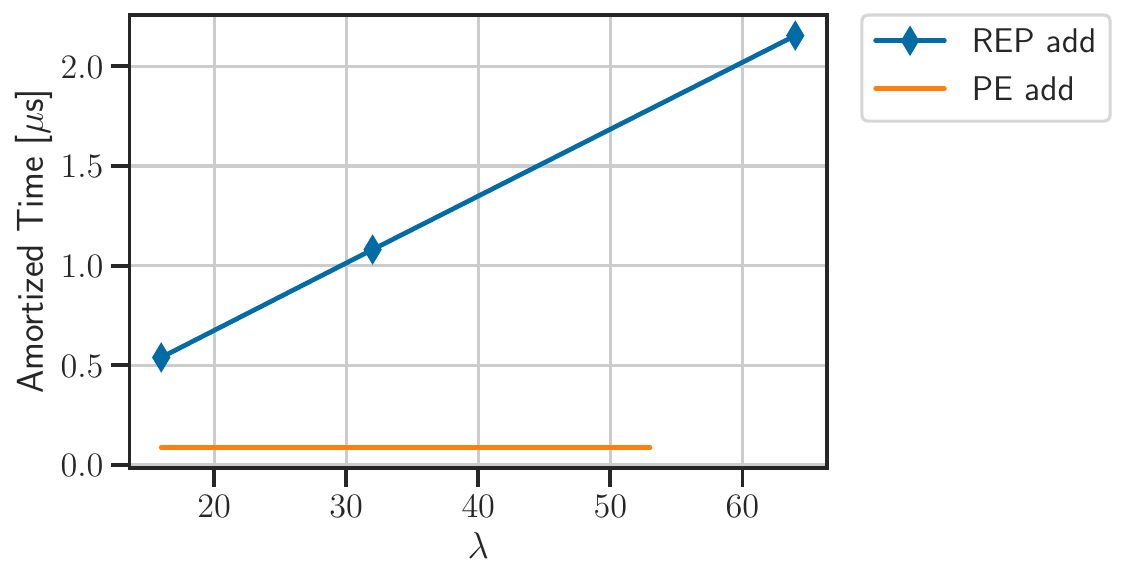}
    \caption{{\small Amortized runtime of the homomorphic addition operation vs. the authenticator security parameter $\lambda$.}}
    \label{fig:lbdAdd}
\end{figure}

We analyze \namecomma' overhead with respect to the authenticator security parameter $\lambda$ for a fixed evaluation circuit. Figure~\ref{fig:lbdAdd} presents our experimental results for the BFV homomorphic addition operation ($\log N{=}14$, $\log q{=}438$) with $\lambda {\in} \{16, 32, 64\}$. We observe that REP induces a linear computation (and communication) overhead with respect to $\lambda$. This is unsurprising since with REP there is an expansion factor of $\lambda$ due to the replication and the challenges (\S\ref{sec:vche1:def}). On the contrary, PE's overhead is constant with respect to $\lambda$ since its security is linked only to the size of the plaintext space (\S\ref{sec:vche2:def}). For appropriate parameterization of the HE scheme, increasing $\lambda$ does not affect the ciphertexts. However, we remind that PE's efficiency is directly linked to the authenticator's growth at every multiplication (\eg \S\ref{sec:eval:inference} and \ref{sec:eval:search}). Moreover, we note that in the Lattigo implementation, the HE circuit constraints the plaintext space and PE can be parameterized to achieve a security level of up to $\lambda {=} 53$ (depending on the ring size and the evaluation circuit). We emphasize that this is only due to the specific implementation and using arbitrary-precision arithmetic would overcome this issue.

\section{Discussion}\label{ap:disc}

We now discuss various aspects related to the authenticators and the developed library \namecomma.

\subsection{Challenge Verification}\label{ap:disc:snarks}

Both authenticators operate in the \textit{designated verifier} setting, \ie they require the challenge computation during the verification procedure by the client (or any holder of the secret key). In some cases, this might be a limitation because the client has to spend resources on this computation (\S\ref{sec:eval}). Nonetheless, these challenges do not leak any information about the initial data (recall that they are computed by the keyed PRF and the identifiers), and they can be offloaded to another entity that does not collude with the server. 
Hence, standard verifiable computation techniques, such as interactive proofs~\cite{goldwasser2015delegating}, SNARKs~\cite{bunz2018bulletproofs} or STARKs~\cite{ben2019aurora}, could be employed to ensure the correct computation on these challenges. With state-of-the-art SNARKs \eg\cite{bunz2018bulletproofs,maller2019sonic,parno2013pinocchio}, this would lead to a sublinear verification complexity in the size of the circuit, albeit with a modification in our system model (\S\ref{sec:pb}). 

\subsection{\namecomma' Overhead in Perspective}\label{ap:disc:perspective}
As observed in Figure~\ref{fig:bench}, compared to the HE baseline, the use of \name introduces an overhead factor between $0.5\times$ and $38\times$ for the client upon offloading and decryption (and verification) and between $0.1\times$ and $3\times$ for the server's evaluation. Although non-negligible, this overhead is actually much smaller than the one already induced by the use of HE. Indeed, compared to a client running the computations directly on its local data, offloading encrypted data is between one and two orders of magnitude more expensive ($10$--$100\times$). At the same time, evaluating the homomorphic circuit on encrypted data at the server is between one and four orders of magnitude slower ($10$--$15,000\times$). For example, in the disease susceptibility use-case (\S\ref{sec:eval:genomics}), the client's offloading work is $600\times$ more expensive than running the computation locally, whereas its decrypting work is $186\times$ more costly. Similarly, the server's operations under encryption are $15,000\times$ more expensive compared to the plaintext evaluation which is much more significant than the overhead created by our encodings. Therefore, \namecomma' overhead for both the client and the server is close to a negligible addition to the cost of HE. Further improvements that lower the overhead of the HE schemes would directly translate to a reduction in \namecomma' overhead as well.

\subsection{Preventing Verification Outcome Leakage}\label{ap:disc:outcome}

\name enables a client to detect with high probability if a server miscomputed a homomorphic evaluation. As described in our threat model (\S\ref{sec:pb}), our work considers a malicious but rational adversary that cheats only if it has a negligible chance of being detected. Such servers are strongly discouraged from tampering with the requested computations in the presence of \namecomma. Yet, a misbehaving server that does not care about being detected could purposefully cheat a homomorphic computation in an attempt to obtain side-channel information about the client's data based on the latter's behavior, \eg its termination of the service, its request to re-compute a computation (that failed the verification check) on the same inputs, etc. To prevent this leakage, as discussed in \S\ref{sec:pb}, application-layer countermeasures can be employed orthogonally to \namecomma. 

We now present an example mechanism that prevents the leakage of the verification outcome to the server by imposing a constant behavior at the client irrespectively of the verification outcome. Informally, the mechanism forces the client to send a fixed number of periodic and independent computation queries to the server. We define a \textit{session} as a logical computation context in which the client performs a maximum number of computation requests to the server at periodic time intervals. The maximum number and frequency of computations are established with an a priori agreement between the client and the server based on the application requirements, \eg the client purchases a subscription for $n$ computation requests to the server which are performed every minute. At the end of a session, the client is expected to disconnect and terminate her interaction with the server. Without loss of generality, we consider that the client engages in a single session with the server; the mechanism can be trivially extended to multiple sessions. 

\descr{Session Setup.} The client establishes a session with the server and they both agree on a homomorphic encryption scheme and the appropriate parameterization (\ie the polynomial rings $\mathcal{R}_q$). Moreover, they agree on the maximum number of computation requests for the session $\textsf{maxNReq}$ and the frequency $\textsf{freq}$ of computation requests based on application requirements; the client sets up a counter $\textsf{count}$ that keeps track of the number of computation requests remaining for the session and sets $\textsf{count}=\textsf{maxNReq}$. The client initializes the session by generating the appropriate key material with $\text{HA.KeyGen}$. This process generates the homomorphic key material (\ie $\mathbf{sk}_{\text{HE}}$, $\mathbf{pk}_{\text{HE}}$, and $\mathbf{sk}_{\text{HE}}$) but also the PRF key $K$ and the encoding's secret (\ie $S$ or $\alpha$).

\descr{Program Generation.} When the client wants to evaluate a function $f(\cdot)$ on her inputs $m_1, \cdots, m_n$ corresponding to the identifiers $\tau_1, \cdots, \tau_n$, she first generates a verifiable program (see \S\ref{sec:prelim:verprog}) $\mathcal{P} = (f(\cdot), \tau_1, \cdots, \tau_n)$. As described in \S\ref{sec:prelim:verprog}, the identifiers $\tau_i$ are strings that uniquely identify the client's private inputs.

\descr{Authentication.} The client authenticates her inputs by calling the homomorphic authenticator's authentication procedure ($\text{HA.Auth}(\cdot)$). The authentication returns a list of ciphertexts encrypting its inputs encoded using the homomorphic encoder.

\descr{Computation Requests.} Every $\textsf{freq}$ timesteps, the client performs a computation request for a specific verifiable program to the server. She sends the list of ciphertexts and the function to be evaluated homomorphically. The server then calls the homomorphic authenticator's evaluation procedure ($\text{HA.Eval}(\cdot)$). After each computation request, the client decreases $\textsf{count}$.

\descr{Computation Verification.} Upon receiving the ciphertexts containing the computation result from the server, the client calls the homomorphic authenticator's verification procedure ($\text{HA.Ver}(\cdot)$). Upon verification failure, the client completely loses her trust in the server. Yet, the client continues performing independent and periodic computation requests to the server until the session termination and she silently discards their results.

\descr{Session Termination.} Once the counter $\textsf{count}$ reaches zero, the client stops using the encoding's secret key and terminates the existing session with the server.\newline

We now briefly analyze this mechanism and we argue that the server does not learn any information about the different verification outcomes. For our analysis, we make the assumption that the verification outcome leakage happens when the client detects a wrongful computation performed by the server and immediately stops using the server for future computations. 

When the client detects a wrongful computation, she does not trust the server anymore. Yet, she still continues the session until its termination, \ie the client performs computation requests until $\textsf{count}{=}0$. All the subsequent requests following the verification failure can be seen as dummy requests for which the client simply silently discards the results. As these computation requests are independent and periodic, the server cannot determine the client's verification bit.\newline

\textit{Proof Sketch.} When $\textsf{count}{=}1$, the client uses the server for a final single computation in the session. As such, the interaction between the two stops directly after the server sends back the resulting ciphertext. Thus, if the server cheats this final computation, it does not gain any information from the verification outcome as no further communication between the client and the server is expected. When $\textsf{count}{>}1$, upon a verification failure, the client keeps sending sequentially independent requests until $\textsf{count}{=}0$ but discards the ciphertexts returned by the server: Those requests can be seen as \textit{dummy} requests. 
Because the requests are periodic and by the semantic security of the homomorphic encryption scheme, from the server's perspective, the client's behavior against the dummy queries is indistinguishable from legitimate queries, \ie the server's view does not change between accepted/rejected computations. 
To support this claim, we propose the following indistinguishability experiment $\textbf{Exp}_{\mathcal{A}, b}^{\text{IND}}$ in which the adversary aims at distinguishing ciphertexts from two databases (Experiment~\ref{expeInd}). Because the computation queries are independent and periodic, they can be seen as parallel events over a database $\text{DB}$ of size $N_{\text{db}}$ comprising plaintexts $\mathbf{m}_{1}, \dots, \mathbf{m}_{N_{\text{db}}}$. 
\begin{figure}[t]
\refstepcounter{expe}\label{expeInd}
\experiment{\textsc{Indistinguishability Experiment} }{
~
\newline$\textbf{Exp}_{\mathcal{A}, b}^{\text{IND}}$:
\begin{itemize}[leftmargin=*, topsep=0pt, itemsep=0ex]
    \item $\text{DB}_0, \text{DB}_1 {\leftarrow} \mathcal{A}(\cdot)$ where $\text{DB}_j{=}\{\mathbf{m}_{j,i}\}^{N_{\text{db}}}_{i=1}$ for $j\in \{0,1\}$.
    \item $b{\in}_{R} \{0,1\}$
    \item \textbf{return} $b'\leftarrow \mathcal{A}^{\mathcal{O}_{\text{Query}}(\cdot)}(\mathbf{pk}, \mathbf{evk})$    
\end{itemize}
}
\end{figure}
This experiment calls a query oracle $\mathcal{O_{\text{Query}}}(\mathcal{P})$ for a verifiable program $\mathcal{P}$. 
\begin{figure}[t]
\refstepcounter{oracle}\label{oracleInd}
\oracle{\textsc{Query Oracle} }{
~
\newline$\mathcal{O_{\text{Query}}}(\mathcal{P})$:
\begin{itemize}[leftmargin=*, topsep=0pt, itemsep=0ex]
    \item parse $\mathcal{P}=(f(\cdot), \tau_1, \cdots, \tau_n)$
    \item \textbf{if} $\textsf{count}=0$ \textbf{return} $\perp$
    \item \textbf{return} $\forall i{\in}[n] \text{ not sent before } \mathbf{c}_i=\text{HE.Enc}(\mathbf{m}_{b,i}; \mathbf{pk}_{\text{HE}})$ and instructions to evaluate $f(\cdot)$
\end{itemize}
}
\end{figure}

The proof proceeds with a sequence of games to show that the adversary has a negligible advantage of winning $\textbf{Exp}_{\mathcal{A}, b}^{\text{IND}}$. Game 0 is the original experiment $\textbf{Exp}_{\mathcal{A}, b}^{\text{IND}}$. In Game 1, we modify Game 0 such that the input ciphertexts are all pre-computed right after the pick of the challenge bit $b$. This mere reordering of the operations does not change the adversary's advantage of winning the game. In Game 2, we replace the pre-computed ciphertext of the inputs by encryptions of zeros. If the adversary could distinguish between Game 1 and Game 2 then it would be able to break the IND-CPA security of the HE scheme. Now, Game 2 is independent of the challenge bit $b$. Thus, the advantage of the adversary winning Game 2 is negligible which concludes the proof.
$\square$
    
The proposed mechanism based on dummy queries successfully prevents the revelation of the verification results under the leakage assumption of a client immediately terminating a session upon the detection of a tampered computation. 
Note that for non-independent computations, the mechanism can be extended to operate in sessions of single computations (\ie $\textsf{count}{=}1$), where the client alters her identity after each session (using an appropriate anonymity mechanism).

\end{appendices}

\end{document}